\documentclass{aa}

\usepackage{graphicx}
\graphicspath{{figs/}}
\usepackage{hyperref}
\usepackage[all]{hypcap}
\usepackage[capitalise]{cleveref}
\usepackage[separate-uncertainty]{siunitx}
\usepackage{placeins}
\usepackage{txfonts}
\usepackage{xcolor}
\usepackage{ulem}
\usepackage{ifthen}

\newboolean{highlightchanges}
\setboolean{highlightchanges}{false} 

\ifthenelse{\boolean{highlightchanges}}{
    \newcommand*{\change}[1]{\textcolor{blue}{\textbf{#1}}}
}{
    \newcommand*{\change}[1]{#1}
}
\begin{document}

   \title{Discovery of a $>$13\,Mpc long X-ray filament between two galaxy clusters beyond three times their virial radii}
   \titlerunning{Discovery of a $>$13\,Mpc long X-ray filament between two galaxy clusters}

   \author{J. Dietl\inst{1}
          \and F. Pacaud\inst{1}
          \and T. H. Reiprich\inst{1}
          \and A. Veronica\inst{1}
          \and K. Migkas\change{\inst{1,2,3}}
          \and C. Spinelli\change{\inst{1}}
          \and K. Dolag\inst{4,5}
          \and B. Seidel\inst{4}
          }

   \institute{Argelander-Institut für Astronomie (AIfA), Universität Bonn, Auf dem H\"ugel 71, 53121 Bonn, Germany \\ \change{e-mail: \texttt{jdietl@astro.uni-bonn.de}}
   \and
   Leiden Observatory, Leiden University, PO Box 9513, NL-2300 RA Leiden, The Netherlands
   \and
   \change{SRON Netherlands Institute for Space Research, Niels Bohrweg 4, NL-2333 CA Leiden, The Netherlands}
   \and
   Universitäts-Sternwarte, Fakultät für Physik, Ludwig-Maximilians-Universität München, Scheinerstr. 1, 81679 München, Germany
   \and
   Max-Planck-Institut für Astrophysik, Karl-Schwarzschild-Straße 1, 85741 Garching, Germany}

   \date{Received ...; accepted ...}

  \abstract
   {A significant fraction of the missing baryons in the local Universe is expected to reside in large-scale filaments that may be observable in soft X-ray emission. Until now, however, very few candidate emission filaments have been found in individual systems, and none beyond three times the virial radius of the clusters at the nodes of these filaments. The new Spectrum Roentgen Gamma (SRG) eROSITA X-ray telescope has a superior response to extended soft X-rays, which makes it ideal for studying low X-ray surface brightness emission of cosmic filaments.}
   {We search for extended X-ray emission between the two nearby galaxy clusters Abell~3667 and Abell~3651, which are separated by a projected transverse distance of ${\sim}\SI{13}{Mpc}$, using data from the SRG/eROSITA All-Sky Survey.}
   {We performed a detailed X-ray image analysis of the region between the two galaxy clusters and conducted a redshift analysis of the sources between them.  We carried out a thorough surface brightness and spectral analysis between the clusters. The analysis was complemented with an X-ray pointed observation from XMM-Newton, infrared 2MASS data, and redshift information from NED.}
   {We discover an emission filament beyond the known radio relic northwest of A3667 and even beyond three times its virial radius. It is smoothly connected to A3651. The X-ray emission in the direction of the filament shows an enhancement of $\SI{30(3)}{\%}$ with a significance of $\SI{11}{\sigma}$. The 2MASS map and redshift analysis show an alignment of galaxies along the filament and make a projection effect unlikely. Taking the redshift progression of galaxies within the filament into account, we estimate its three-dimensional length to be in the range of $\SI{25}{\mega pc} - \SI{32}{\mega pc}$. The surface brightness analysis in combination with the temperature $T=(\num{0.91}_{-0.11}^{+0.07})\,\mathrm{keV}$ and metallicity $Z=(\num{0.10}_{-0.08}^{+0.05})\,\mathrm{Z_{\odot}}$ from the spectral analysis leads to estimates of a total flux, gas mass, and central baryon overdensity of $F_\mathrm{X}= \num{7.4(12)e-12}\,\mathrm{erg\,s^{-1}\,cm^{-2}}$, $M_\mathrm{g}=(2.7^{+1.4}_{-0.8})\times 10^{14}\,\mathrm{M_\odot}$ and $\delta_0=215^{+86}_{-50}$.}
  {}

   \keywords{Galaxies: clusters: individual: Abell 3667, Abell 3651 – X-rays: galaxies: clusters - intergalactic medium
               }

   \maketitle
%

\section{Introduction}
\label{sec:introduction}
Galaxy clusters offer a unique opportunity to gain deep insights into the formation of the large-scale structure (LSS) of the Universe. In the framework of hierarchical structure formation, which is a key ingredient of the concordance ${\Lambda}$CDM model, galaxy clusters are located at the nodes of a cosmic web structure that is connected by filaments and sheets of galaxies and gas (e.g., \citealt{Bond}). This prediction is confirmed by observations, where galaxy filaments have been known for decades. It is also a well-established fact that the amount of baryonic matter that is directly detected in the Universe today does not match the expectation from primordial Big Bang nucleosynthesis and observations of the cosmic microwave background. The question of where the remaining fraction of baryons ($\approx\num{30}$--$\SI{40}{\percent}$) may be is called the missing baryon problem (e.g., \citealt{missingb}).

Simulations of the LSS have revealed that a substantial fraction of the missing baryons could be located in the so-called warm-hot intergalactic medium (WHIM; e.g., \citealt{missingsim,Martizzi}), namely hot intergalactic gas in the temperature range of $\num{e5}$--$\SI{e7}{\kelvin}$. Although absorption studies are usually considered to be more sensitive, the high-temperature fraction of this WHIM may be observable in emission in the soft X-ray range. \change{The most promising strategy for such detections of the WHIM is to search near galaxy clusters and in their outskirts, for example, Abell 2744 (e.g., \citealt{2744}) or Abell 1750 (e.g., \citealt{1750}), or for the presumably shorter filaments between galaxy clusters, which are expected to be hotter and denser, for example, Abell 222/223 (e.g.,\citealt{222_223}), Abell 399/401 (e.g., \citealt{399_401_2, 399_401}), Abell 98N/S (e.g., \citealt{98_N_S, Abell98_Sarkar}), or Abell 3391/95 (e.g., \citealt{Abell9195,3391_95}). However, even so, their density remains low, and the detection of the weak X-ray emission from these filaments remains difficult (e.g., \citealt{missingb}). In particular, no individual intergalactic filaments with significant emission extending beyond the projected $3R_{200}$ radius have been observed so far.}

In this work, we search for enhanced X-ray emission between the two galaxy clusters Abell~3667 (A3667) and Abell~3651 (A3651) using the unique response to extended soft X-rays of the extended ROentgen Survey with an Imaging Telescope Array (eROSITA). We use data from its all-sky survey~1-4 (eRASS:4) to do this.

\begin{figure}[htbp]
    \centering
    \includegraphics[width=\hsize]{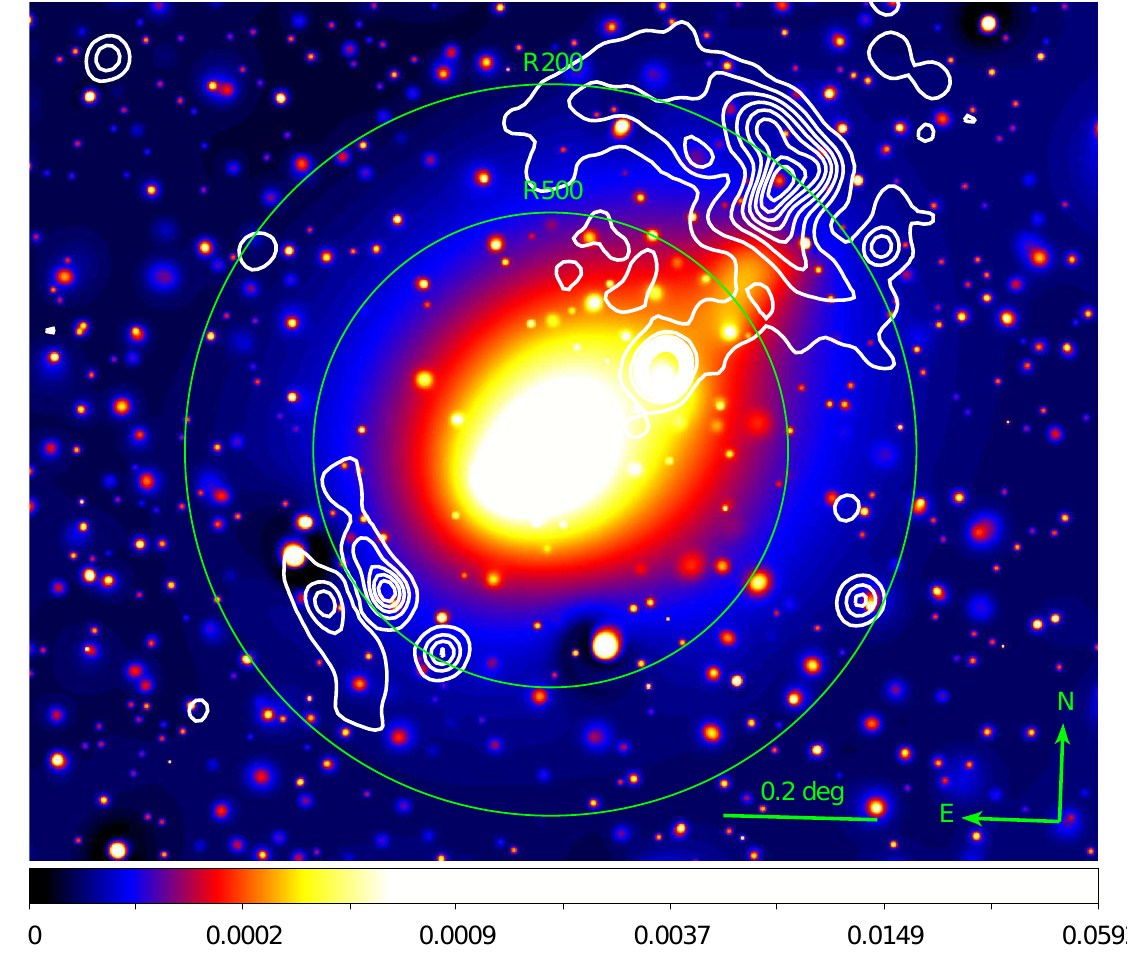}
    \caption{eROSITA X-ray image of A3667 in the soft X-ray band with radio contours (\num{170}–\SI{231}{MHz}) from the GaLactic and Extragalactic All-Sky MWA Survey \citep{GLEAM}. The well-known radio relics and the substructure in the northwest are clearly visible. The image was created and corrected using the methods described in Sect. \ref{subsec:eROSITA_DR}. The color bar is in units of counts per second (as in all following images). In addition, the characteristic radii $R_{500}$ and $R_{200}$ are displayed.}
    \label{Fig:A3667}
\end{figure}

The galaxy cluster A3667 (see \cref{Fig:A3667}) is one of the bright nearby galaxy clusters located in the southern hemisphere (e.g., \citealt{BrightC}). It features well-known radio relics in the northwest and southeast (e.g., \citealt{RadioRelics, Gasperin_A3667}). The northwest radio relic is one of the strongest ever observed (e.g., \citealt{PowerfulRadio}). Furthermore, evidence of a significant substructure in the northwest of A3667 was found (e.g., \citealt{substruc}), as was a prominent cold front southeast of the cluster (e.g., \citealt{coldfront}).
The galaxy cluster A3651, on the other hand, has not been analyzed in great depth. It is also a strong X-ray emitter and seems to form a pair of galaxy clusters together with A3667 \citep{A3651}. The two clusters feature similar redshifts and are ${\sim}\SI{3.34}{\degree}$ apart on the sky. The properties of both clusters are listed in \cref{tab:clusterprops}. The values are taken from the Meta-Catalogue of X-ray Detected Clusters of Galaxies (MCXC; \citealt{MCXC}).
The table shows the redshift $z$, the right ascension (RA) and declination (DEC) of the coordinates in the fk5 system, and the characteristic radii $R_{500}$ and $3R_{200}$, between which the cluster outskirts can be roughly defined \citep{OutskirtsPaper}.  The system of the two galaxy clusters is shown in \cref{Fig:SystemOverview}.
\begin{table}
   \caption{General properties of A3667 and A3651: Redshift, position in fk5, and radii}
      \label{tab:clusterprops}
  $$ 
      \begin{array}{lll}
         \hline
         \noalign{\smallskip}
         \textnormal{Cluster} & \textnormal{A3667} & \textnormal{A3651} \\
         \noalign{\smallskip}
         \hline
         \noalign{\smallskip}
         z  & \num{0.0556} & \num{0.06} \\
         \textnormal{RA} / \si{\degree} & \num{303.127} & \num{298.069} \\
         \textnormal{DEC} / \si{\degree} & \num{-56.832} & \num{-55.062} \\
         R_{500} / \si{arcmin}  & \num{18.503} & \num{11.334} \\
         3R_{200} / \si{arcmin} & \num{85.399} & \num{52.313} \\
         \noalign{\smallskip}
         \hline
      \end{array}
  $$ 
\end{table}
The assumed cosmology in this work is a flat $\Lambda$CDM, with the cosmological parameters $h = \num{0.7}$, $\Omega_{\text{m}} = \num{0.3}$, and $\Omega_{\Lambda} = \num{0.7}$. At the redshift of A3667, $\SI{1}{arcsec}$ corresponds to $\SI{1.08}{kpc}$.

\section{Data reduction}
\subsection{eROSITA}
\label{subsec:eROSITA_DR}
We used data from the eROSITA all-sky survey~1-4 (eRASS:4) of the sky tiles 304147, 298144, 299147, and 303144 using the eROSITA pipeline-processing version~c946. The data from all seven telescope modules (TMs) were used. The data reduction was performed with \texttt{HEASoft} \footnote{\href{https://heasarc.gsfc.nasa.gov/docs/software/heasoft/}{https://heasarc.gsfc.nasa.gov/docs/software/heasoft/}} version~6.12 and the eROSITA Science Analysis Software System \texttt{eSASS} \change{(\citealt{Brunner_2022, Merloni_2024})} \change{(eSASS4DR1\footnote{\href{https://erosita.mpe.mpg.de/dr1/eSASS4DR1/}{https://erosita.mpe.mpg.de/dr1/eSASS4DR1/}})}.

For our analysis, we created images in the $\num{0.3}$--$\SI{2.0}{\kilo\electronvolt}$ band. In the case of TMs 5 and 7, which are not equipped with an on-chip filter, we restricted the band to $\num{0.8}$--$\SI{2.0}{\kilo\electronvolt}$, however. 
This was done to mitigate the influence of leaked optical light from the Sun, which can enter through a gap in the detector shielding, and in the absence of an on-chip filter, pile up in each CCD pixel and mistakenly be detected as soft X-ray photons \citep{eROSITA}.

All the data reduction steps mentioned below follow the procedure used in \cite{Abell9195}, where more detailed information about the data reduction processes can be found.

\begin{figure*}[htbp]
    \centering
    \includegraphics[width=0.9\hsize]{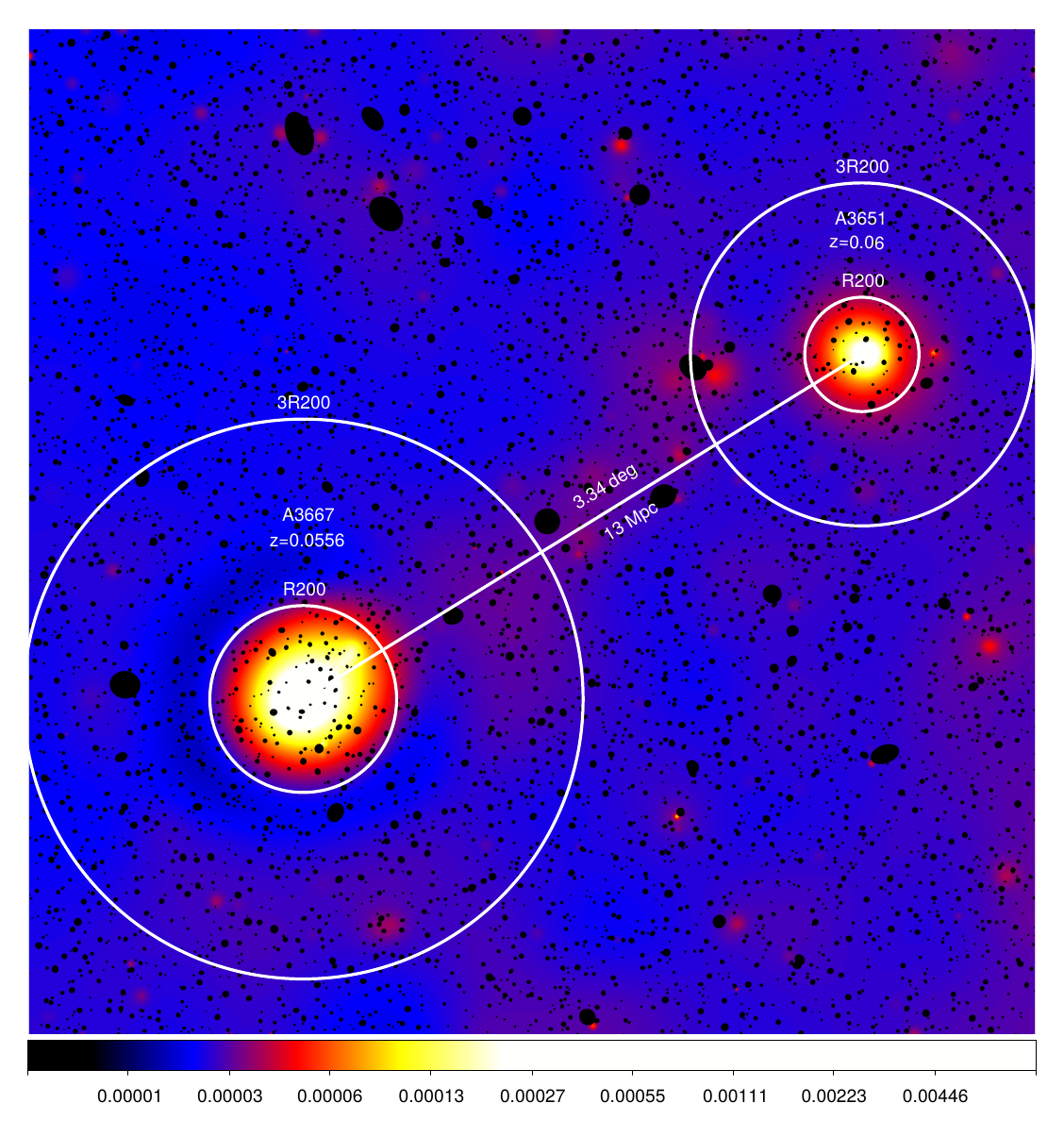}
    \caption{Data-reduced and wavelet-filtered eRASS:4 X-ray image of the Abell 3667 -- Abell 3651 system in the $\num{0.3}$--$\SI{2.0}{\kilo\electronvolt}$ band. Additional information for both clusters is shown, as is the projected length from one cluster center to the other.}
    \label{Fig:SystemOverview}
\end{figure*}

After combining the skytiles, we extracted and visually inspected the light curves to ensure that the data were not affected by any particle flare. A particle-induced background (PIB) map was then created using observation data between $\num{6}$--$\SI{9}{\kilo\electronvolt}$, converted into the lower-energy band using the hardness ratios observed in the eROSITA filter-wheel-closed (FWC)\footnote{\href{https://erosita.mpe.mpg.de/edr/eROSITAObservations/EDRFWC/}{https://erosita.mpe.mpg.de/edr/eROSITAObservations/EDRFWC/}} data and distributed across the image following the unvignetted exposure map. This PIB map was subtracted from all images and surface brightness products. Our exposure correction accounts for different exposure times and vignetting throughout the image, but also corrects for the use of different energy bands for the different TMs. The raw exposure maps were modulated by an $N_{\text{HI}}$ absorption correction using data from the HI4PI all-sky survey \citep{HI4PI} in order to correct for the different X-ray absorption in different regions of the sky due to interstellar medium in the line of sight. The $N_{\text{HI}}$ correction factor map was calculated using simulated expected count rate values. This correction factor map shows a variation of less than $\SI{5}{\percent}$.
These steps were repeated for all four surveys, and the surveys were combined afterward.
For the purpose of the image analysis, for instance,~for extracting surface brightness profiles, the point sources have to be removed. In order to do this, a wavelet filtering was applied, and the software \texttt{Source Extractor} (SExtractor, \citealt{SExtractor}) was used to automatically detect and extract point sources, following the procedure described in \citet{Pacaud_2006,2018A&A...619A.162X}. In addition to this procedure, a few additional apparent point sources that were not detected were manually added to the list, and detected apparent extended sources were excluded from the list. A final wavelet-filtered map was produced after excluding the point sources with a cheesemask in order to reduce wavelet-ringing artifacts around the bright compact sources. The final image to which all correction steps were applied is shown in \cref{Fig:SystemOverview}. Additional information about the cluster names, redshifts, and characteristic radii from the MCXC is displayed.

From now on, we refer to the data from which the point sources and particle-induced background was subtracted and that were corrected for $N_{\text{HI}}$ absorption and exposure as the corrected data. All eROSITA images show the count rate of the combined 7~TMs in the energy band $\num{0.3}$--$\SI{2.0}{\kilo\electronvolt}$ in units of counts per second.

\subsection{XMM-Newton}
In the course of the analysis, a projection effect of the extended foreground object S0840 must be ruled out. Therefore, an additional analysis of S0840 was performed using publicly available \textit{XMM-Newton} data with Obs-ID ${0673180401}$.

\begin{figure}[htbp]
    \centering
    \includegraphics[width=\hsize]{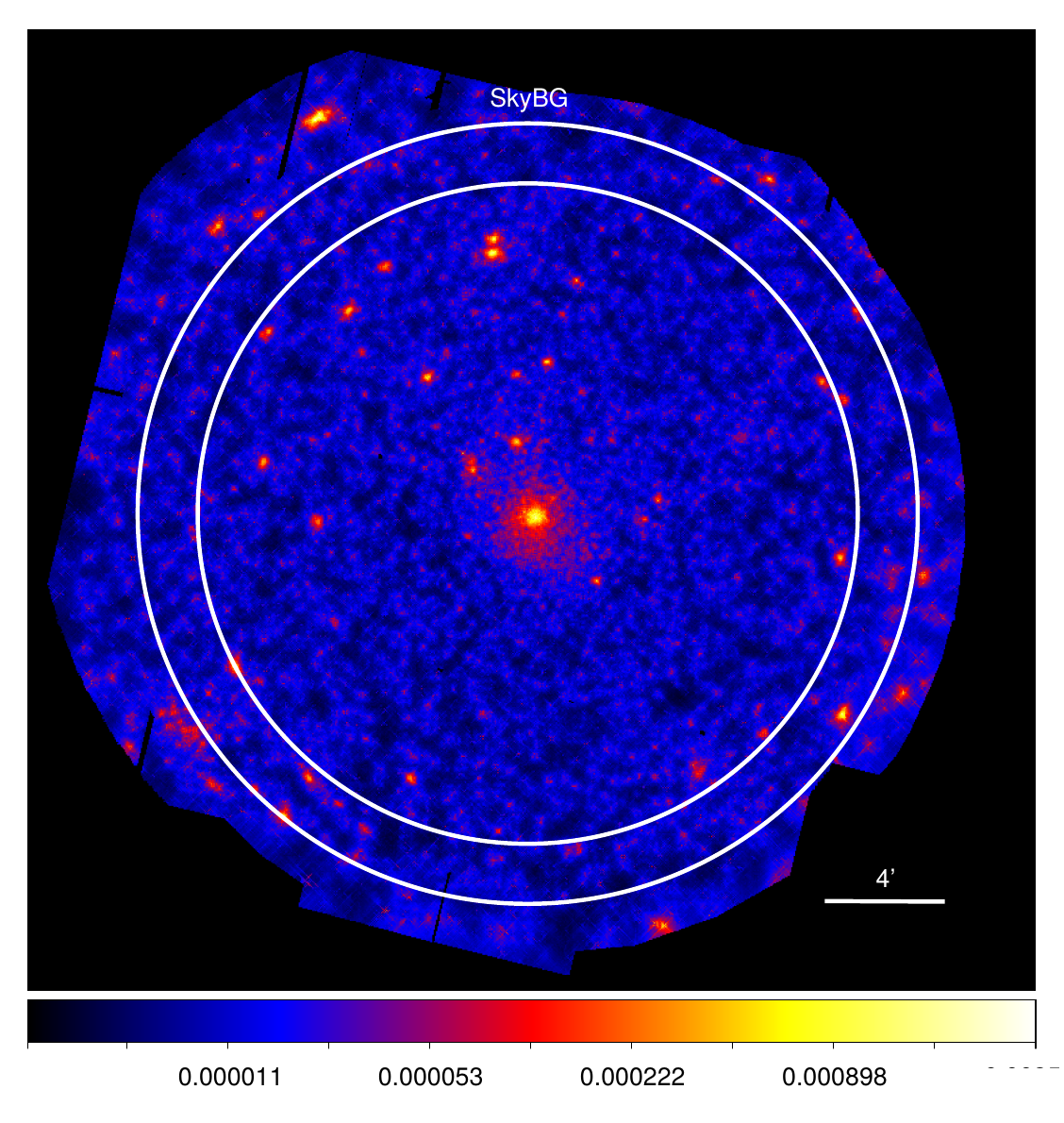}
    \caption{Data-reduced \textit{XMM-Newton} image of Abell S0840 in the energy band $\num{0.5}$--$\SI{2.0}{\kilo\electronvolt}$, smoothed with a Gaussian kernel radius of 5\,pixels. The area between the circles at $\SI{11}{arcmin}$ and $\SI{13}{arcmin}$ indicates the region used for the sky background in the analysis.}
    \label{Fig:S0840}
\end{figure}

The data reduction procedure followed the steps explained in \citet{Northern_Clump}. The Science Analysis Software (SAS) version 20.0.0 was used for the analysis.
The first step involved filtering for soft proton flares using a 3$\sigma$ clipping procedure on the light curves in the $\num{0.3}$--$\SI{10.0}{\kilo\electronvolt}$ energy band.
Furthermore, EPIC-MOS CCDs with an unusually low hardness ratio and a high total background rate, referred to as being in ``anomalous state'' \citep{QPB}, were excluded from the analysis.
The PIB was modeled for each CCD individually using rescaled FWC data and was subtracted from observations, together with a model of the EPIC-PN out-of-time events.
For the image creation, data from the three \textit{XMM-Newton} detectors MOS1, MOS2, and PN were combined in the $\num{0.5}$--$\SI{2.0}{\kilo\electronvolt}$ energy range, and point sources were removed following the method used in Sect. \ref{subsec:eROSITA_DR} for eROSITA.
For the image shown in \cref{Fig:S0840}, a combined exposure map, weighted according to the effective area of each camera, was used to correct for the different exposure times throughout the image.

\section{Analysis and results}
\label{sec:analysis}
As shown in \cref{Fig:SystemOverview}, both clusters clearly emit in the X-rays up to their respective $R_{200}$. Around A3651, enhanced emission can be seen even beyond $R_{200}$. The cluster itself appears to be highly spherical. The shape of A3667 is elliptical, and the emission extends southeast and northwest. Particularly the northwestern extension that was discussed in the literature (e.g., \citealt{substruc}) is clearly visible (see also \cref{Fig:A3667}).

Furthermore, in \cref{Fig:SystemOverview} the cluster emission does not seem to extend beyond $R_{200}$ (except for the northwest), and in the northeast, there is even less emission than in all other parts of the image. This is not a real feature of the cluster, but an image artifact from the wavelet filtering. More information and analyses about the wavelet artifact can be found in \cref{sec:app_artifact}.

Between the two clusters, a filament structure can be seen that extends from northwest of A3667 to east of A3651. It shows less emission than the clusters themselves, but more emission than other parts of the image that correspond to the X-ray sky background. As shown in \cref{Fig:SystemOverview}, the enhanced emission is beyond the $3R_{200}$ radii of both clusters, where the end of cluster outskirts can be roughly defined. Furthermore, when we assume $R_{\textnormal{virial}} \approx \num{1.36}R_{200}$ \citep{OutskirtsPaper}, the emission extends even beyond three times the virial radii of both clusters. This indicates that the emission does not originate from the intracluster medium, but possibly from a WHIM filament. The projected cluster separation is $\SI{3.34}{\degree}$ from one cluster center to the other. This corresponds to a transverse length of ${\sim}\SI{13}{\mega pc}$ (using $z=\num{0.0556}$, i.e., the redshift of A3667), which provides us with a lower limit for the full extent of the structure, neglecting its size along the line of sight. A more detailed discussion of the filament length is presented in Sect. \ref{sec:filament_length}. The following analysis focuses on this filament structure that connects A3667 and A3651.

\subsection{Surface brightness analysis between the clusters}
\label{sec:filament_SB}
\begin{figure}[htbp]
    \centering
    \includegraphics[width=\hsize]{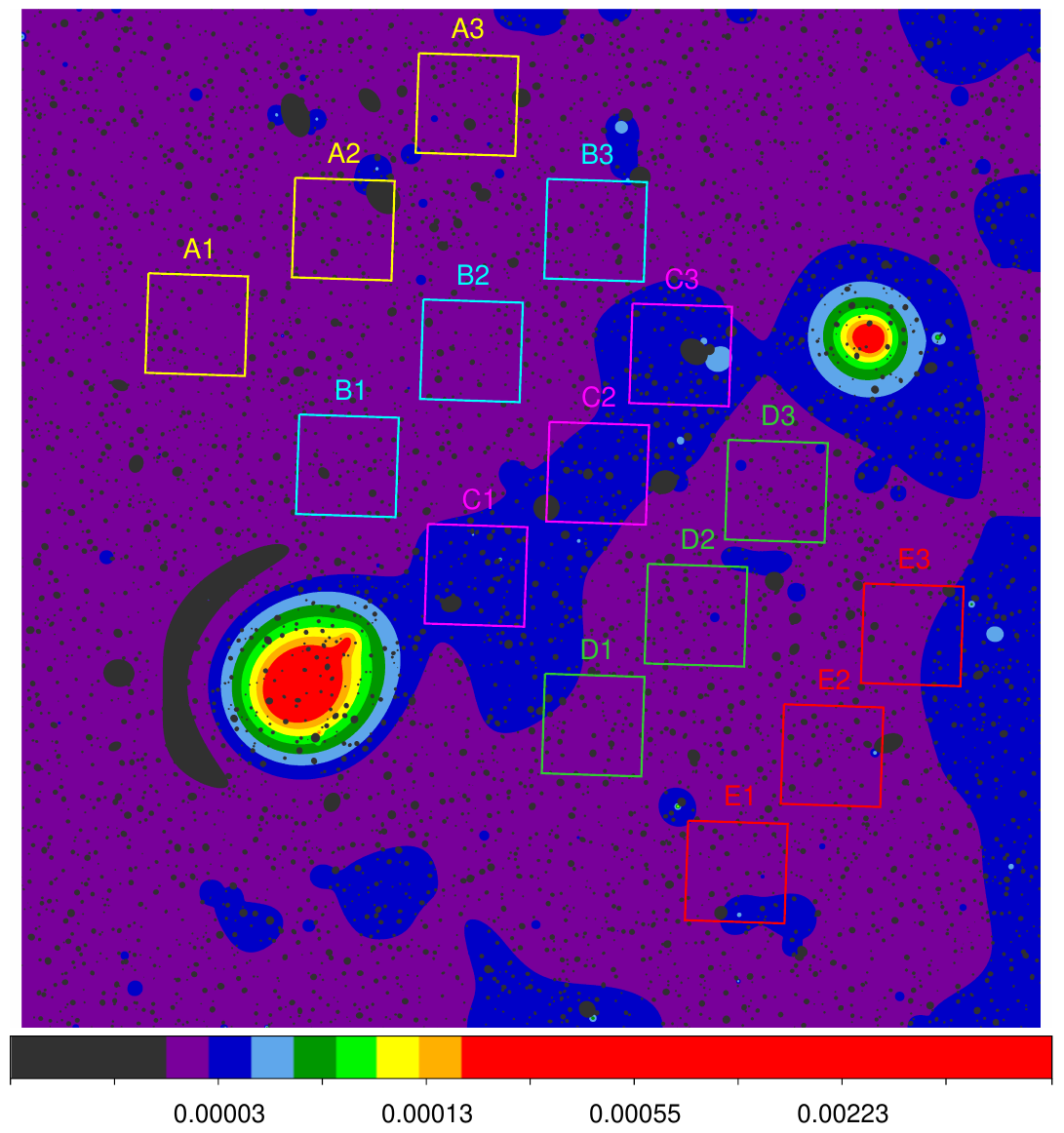}
    \includegraphics[width=\hsize]{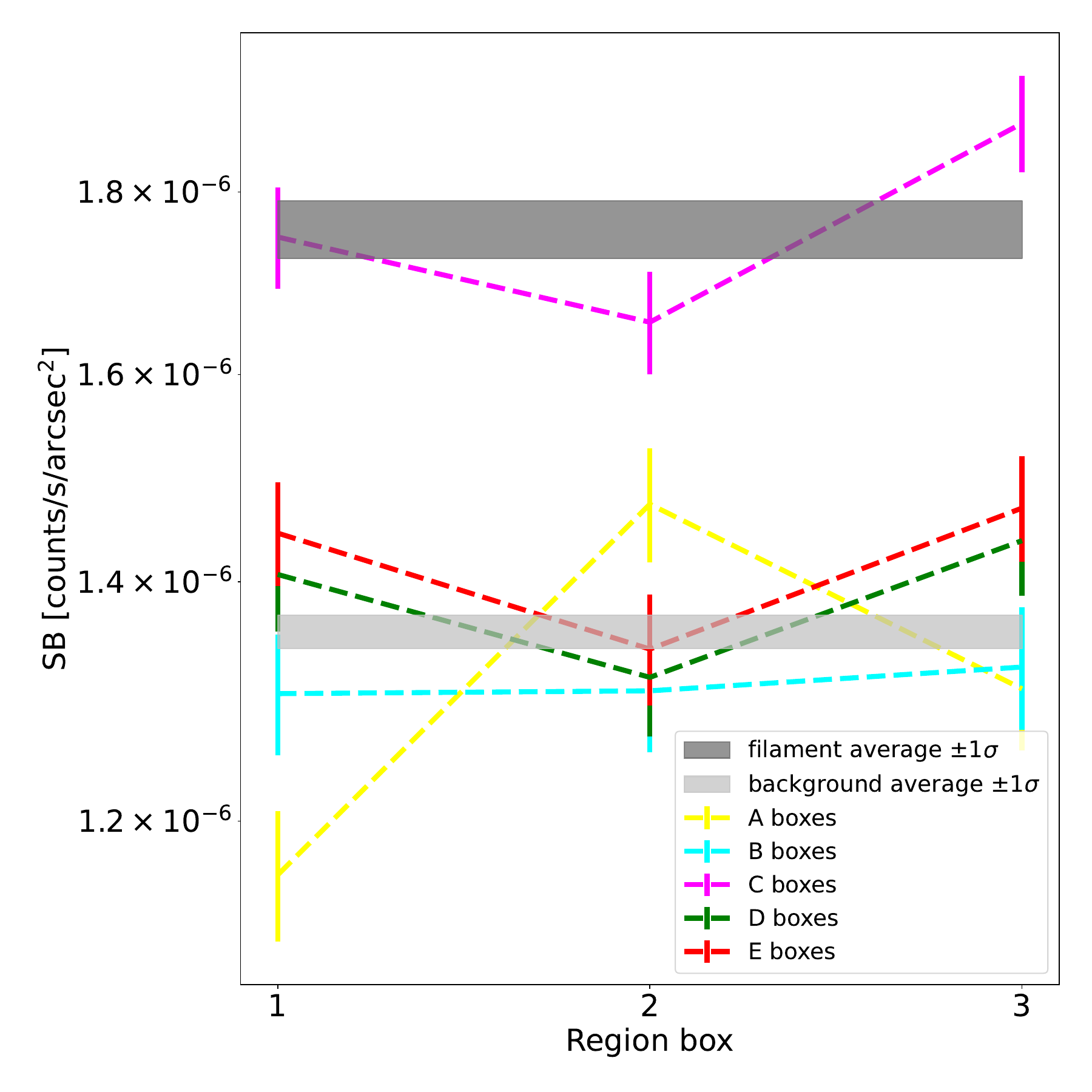}
    \caption{Surface brightness analysis of filament and background. Top: Box placement in the wavelet-filtered image. Bottom: Results of the SB analysis from the top boxes. The dashed lines between the data points are only for the purpose of better visualization and have no physical meaning. The dark gray band shows the weighted arithmetic mean with the $\SI{1}{\sigma}$ interval of all magenta boxes, and the light gray band shows the weighted arithmetic mean with the $\SI{1}{\sigma}$ interval of all background boxes.}
    \label{Fig:SB_boxes}
\end{figure}

A surface brightness (SB) analysis of the region between the two clusters was carried out to quantify the emission in the apparent filament. In order to do this, the SB was calculated in three $\SI{0.5}{\degree}\times\SI{0.5}{\degree}$ boxes along the apparent filament, using the corrected data. To compare the results with other regions of the image, four further groups with three $\SI{0.5}{\degree}\times\SI{0.5}{\degree}$ boxes each were placed parallel to the first three boxes. These boxes were used as reference regions to quantify the emission enhancement between the two clusters as compared to the background. The positions of all boxes are shown in the top panel of \cref{Fig:SB_boxes}.
The SB was calculated via
\begin{equation}
        \label{eqn:SB_formula}
        \textnormal{SB} = \frac{c_\text{image}-c_\text{PIB}}{c_\text{expmap}\cdot A} \ ,
\end{equation}
where $c_{\text{image}}$ and $c_{\text{PIB}}$ are the number of counts in the photon image and the PIB map, respectively. $c_{\text{expmap}}$ is the average exposure per pixel in the source region. $A$ is the area of the box, from which the excised area of the removed point sources was subtracted. The errors on $c$ were estimated from the Poisson standard deviation $\sqrt{c}$, and the final errors were then computed with the Gaussian law of error propagation.

The results are shown in \cref{Fig:SB_boxes} (bottom). The colors in the plot match the colors of the corresponding boxes. The chosen background boxes show some
variation throughout the image, but all have considerably lower values than the values of the C (magenta) boxes in the filament. Particularly the B (cyan) and D (green) boxes directly next to the C (magenta) ones show very little intrinsic fluctuations.
Since the background emission is considerably lower than the filament emission and in order to quantify the filament emission compared to the background emission, the weighted mean of all three C (magenta) boxes was taken for the filament emission~$\overline{\textnormal{SB}}_{\text{filament}}$, and the weighted mean of all background boxes was taken for the background emission~$\overline{\textnormal{SB}}_{\text{bkg}}$. The formulae used for the weighted mean~$\overline{\textnormal{SB}}$ and its error~$\Delta\overline{\textnormal{SB}}$ were
\begin{equation}
        \overline{\textnormal{SB}} = \frac{\sum_{i=1}^{N} \frac{\textnormal{SB}_i}{(\Delta \textnormal{SB}_i)^2}}{\sum_{i=1}^{N} \frac{1}{(\Delta \textnormal{SB}_i)^2}} \quad \text{ and } \quad \Delta\overline{\textnormal{SB}} = \sqrt{\frac{1}{\sum_{i=1}^{N} \frac{1}{(\Delta \textnormal{SB}_i)^2}}}\, ,
\end{equation}
where $\textnormal{SB}_i$ are the SB values of the individual boxes with their errors $\Delta \textnormal{SB}_i$. The relative difference between filament emission and background emission can now be determined with
\begin{equation}
        \frac{\overline{\textnormal{SB}}_\text{filament}-\overline{\textnormal{SB}}_\text{bkg}}{\overline{\textnormal{SB}}_\text{bkg}}\cdot\SI{100}{\percent} = \SI{30(3)}{\percent}
\end{equation}
with a significance of
\begin{equation}
        \frac{\overline{\textnormal{SB}}_\text{filament}-\overline{\textnormal{SB}}_\text{bkg}}{\sqrt{\left(\Delta\overline{\textnormal{SB}}_\text{filament}\right)^2 + \left(\Delta\overline{\textnormal{SB}}_\text{bkg}\right)^2}} = \SI{11(1)}{\sigma}\,.
\end{equation}
By slightly changing the position of the background boxes (e.g., by moving the E (red) boxes more toward or away from the blue structure in the bottom right corner of \cref{Fig:SB_boxes} top), these values only vary within their error margins. We give an error for both the relative difference and the significance to investigate exactly this.
These SB results are discussed in detail in Sect. \ref{sec:discussion}.

A second way of showing the enhanced emission between the two clusters compared to the background is to extract a transverse SB profile of the filament. This is done in Sect. \ref{sec:filament_gas} \cref{fig:transverse_SB} for the analysis of the gas properties.

\subsection{Sources between the clusters}
\subsubsection{Abell S0840}
Inside the filament region, an extended source with a radius of $\SI{4}{arcmin}$ was excised. It is visible in the top panel of \cref{Fig:SB_boxes} as a black circular disk located in the bottom left corner of box C2. At this location, according to the MCXC, a third galaxy cluster is present in the foreground: Abell~S0840, with a redshift $z=\num{0.0148}$ and $R_{500} = \SI{15.6}{arcmin}$ \citep{MCXC}. With its $R_{500}$, it is expected to have a comparable projected size to the other two clusters and might cause a large fraction of the enhanced emission between A3667 and A3651. Because it is in the foreground, however, this emission would not originate from a filament between the two clusters. However, the following analysis shows that the emission of the object S0840 does not contaminate the filament analysis.

In order to show this, we analyzed publicly available \textit{XMM-Newton} data of the S0840 system. While the eROSITA all-sky survey is well suited for large-scale images such as \cref{Fig:SystemOverview}, the \textit{XMM-Newton} data are better suited for this individual object because the \textit{XMM-Newton} pointed observation exposure time of ${\sim}\SI{50}{ks}$ results in better photon statistics than the eROSITA survey exposure time of ${\sim}\SI{2}{ks}$. Furthermore, \textit{XMM-Newton} provides a better spatial resolution with its on-axis PSF of ${\sim}\SI{15}{arcsec}$ \citep{XMM_PSF} compared to the eROSITA survey-averaged PSF of ${\sim}\SI{30}{arcsec}$ \citep{Merloni_2024}.

\begin{figure}[htbp]
    \centering
    \includegraphics[width=\hsize]{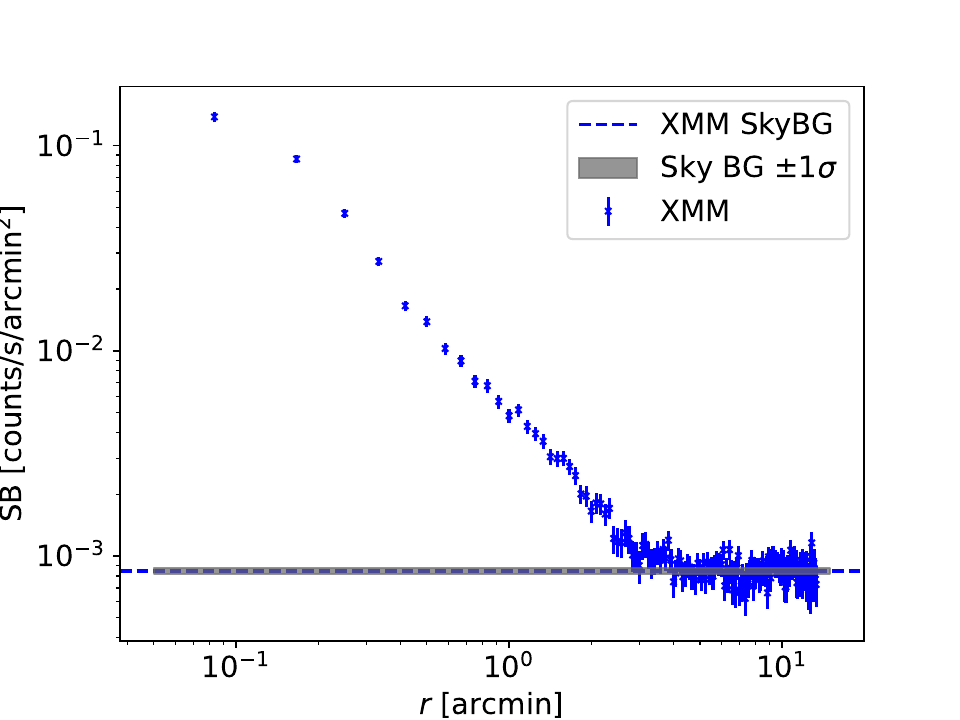}
    \caption{Surface brightness profile of Abell S0840 centered on the brightest pixel. The binning is $\SI{5}{arcsec}$.}
    \label{Fig:SB_S0840}
\end{figure}

From \cref{Fig:S0840} it is already clear that the X-ray emission from S0840 is considerably lower than expected. The outer circle is at $\SI{13}{arcmin} = \num{0.83}R_{500}$. Strong X-ray emission can only be seen in the center. Toward $R_{500}$, no X-ray emission is visible at all. To estimate the radius up to which significant emission can be seen, an SB profile was extracted from the \textit{XMM-Newton} data. In order to do this, we removed point sources from the \textit{XMM-Newton} data and calculated the SB in annuli with a size of $\SI{5}{arcsec}$ around the brightest pixel (RA~$=\SI{300.861}{\degree}$, DEC~$=\SI{-55.947}{\degree}$) using \cref{eqn:SB_formula}. Figure \ref{Fig:SB_S0840} shows the resulting SB profile. The sky background was estimated from the annulus in \cref{Fig:S0840} and is displayed with its uncertainty in the plot. The surface brightness decreases very rapidly from the center toward larger radii and quickly fades into background emission. The estimated sky background matches the surface brightness well at large radii. This plot clearly shows that significant X-ray emission can only be detected up to a radius of $\SI{4}{arcmin}$ at most. Therefore, it was decided to just excise a region of $\SI{4}{arcmin}$ for the SB analysis of the filament. A more detailed analysis of the X-ray flux of Abell~S0840, showing that the flux beyond these $\SI{4}{arcmin}$ is negligible, is carried out in \cref{sec:app_S0840}.

\subsubsection{2MASS}

\begin{figure*}[htbp]
    \centering
    \includegraphics[width=0.9\hsize]{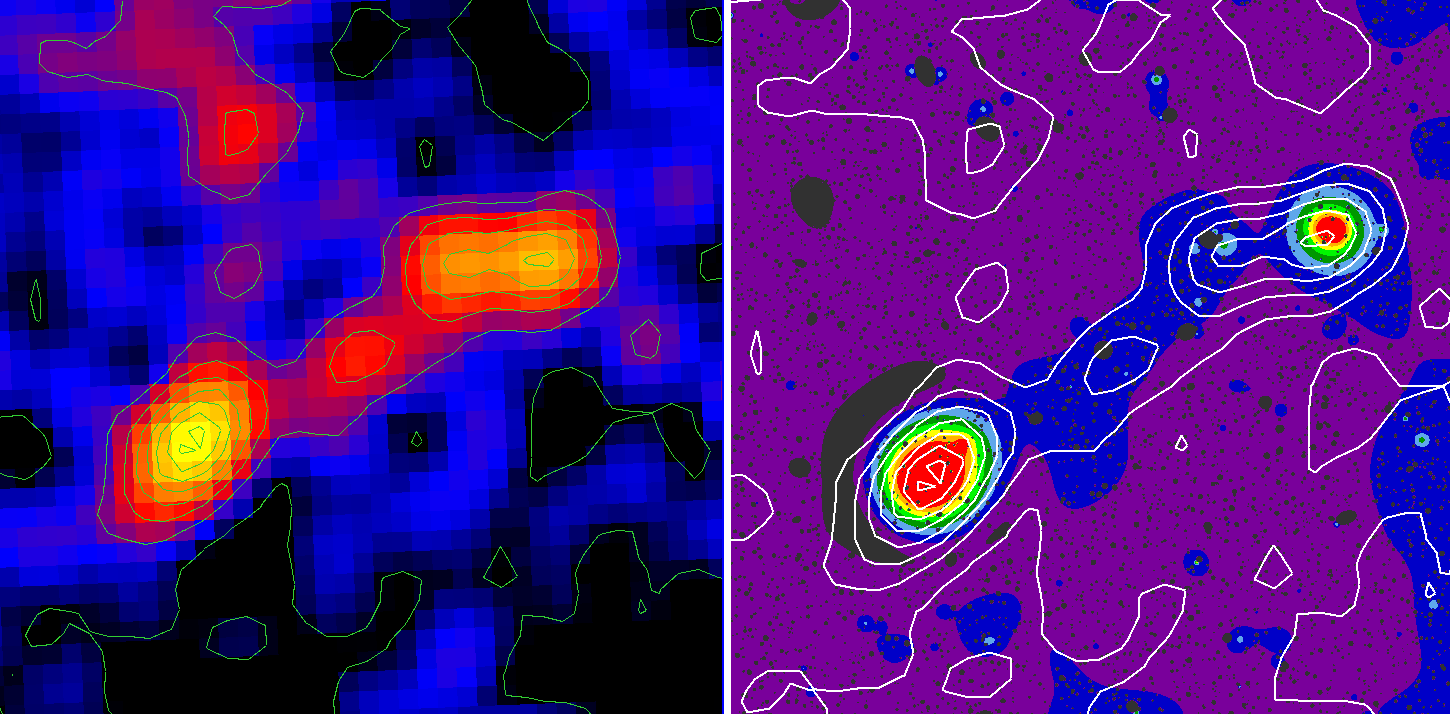}
    \caption{Comparison of galaxy density and X-ray emission. Left: Cut and reprojected 2MASS map from \cite{2MASScat} and \cite{2MASSmap} smoothed with a Gaussian kernel with a radius of 3~pixels and with green contours (contour levels are $\num{0.18}$, $\num{0.73}$, $\num{1.27}$, $\num{1.82}$, $\num{2.36}$, $\num{2.91}$, $\num{3.45}$ and $\num{4.00}$). The brighter regions can be interpreted as regions with a galaxy overdensity compared to the darker regions. Right: Data-reduced wavelet-filtered X-ray image overlaid with the contours from the left image. The brighter regions are regions with enhanced cluster emission compared to the darker regions.}
    \label{Fig:2MASS_both}
\end{figure*}

At the position of a galaxy cluster, an overdensity of galaxies is clearly expected. The same is true for a connecting filament between galaxy clusters. The \textit{Two Micron All Sky Survey (2MASS)} is a survey in the infrared that has, among other things, resulted in an extended source catalog consisting of $\SI{1.6}{million}$ galaxies (\citealt{2MASScat}, \citealt{2MASS}). An all-sky map with a resolution of $\SI{10}{arcmin}$ from \cite{2MASSmap} was cut and reprojected to the image section used in this analysis and was used to estimate the galaxy overdensity in the regions between and around the clusters. The two clusters A3667 and A3651 are clearly visible in the 2MASS map (\cref{Fig:2MASS_both} left), and a bridge can be seen directly between the clusters. By overlaying the contours from the 2MASS map with an X-ray image (\cref{Fig:2MASS_both} right), a correlation between the 2MASS contours and the X-ray filament structure can be seen. The extension of the contours east of A3651 is particularly interesting. At the same position, many point sources were removed, which means that another galaxy group is probably present. This is further elaborated in Sect. \ref{sec:discussion}. The region of Abell S0840 is also of interest, where  a galaxy overdensity can indeed be seen, but it is not as large as the overdensity in the two other galaxy clusters. Additionally, it is unclear to what extent the galaxies in this overdensity are part of S0840 or part of the connecting filament between A3667 and A3651. Consequently, a redshift analysis of these galaxies was performed and is reported in the next section.

\subsubsection{Redshift and spatial analysis of the galaxies}
\label{sec:z_analysis}
\begin{figure}
   \centering
   \includegraphics[width=\hsize]{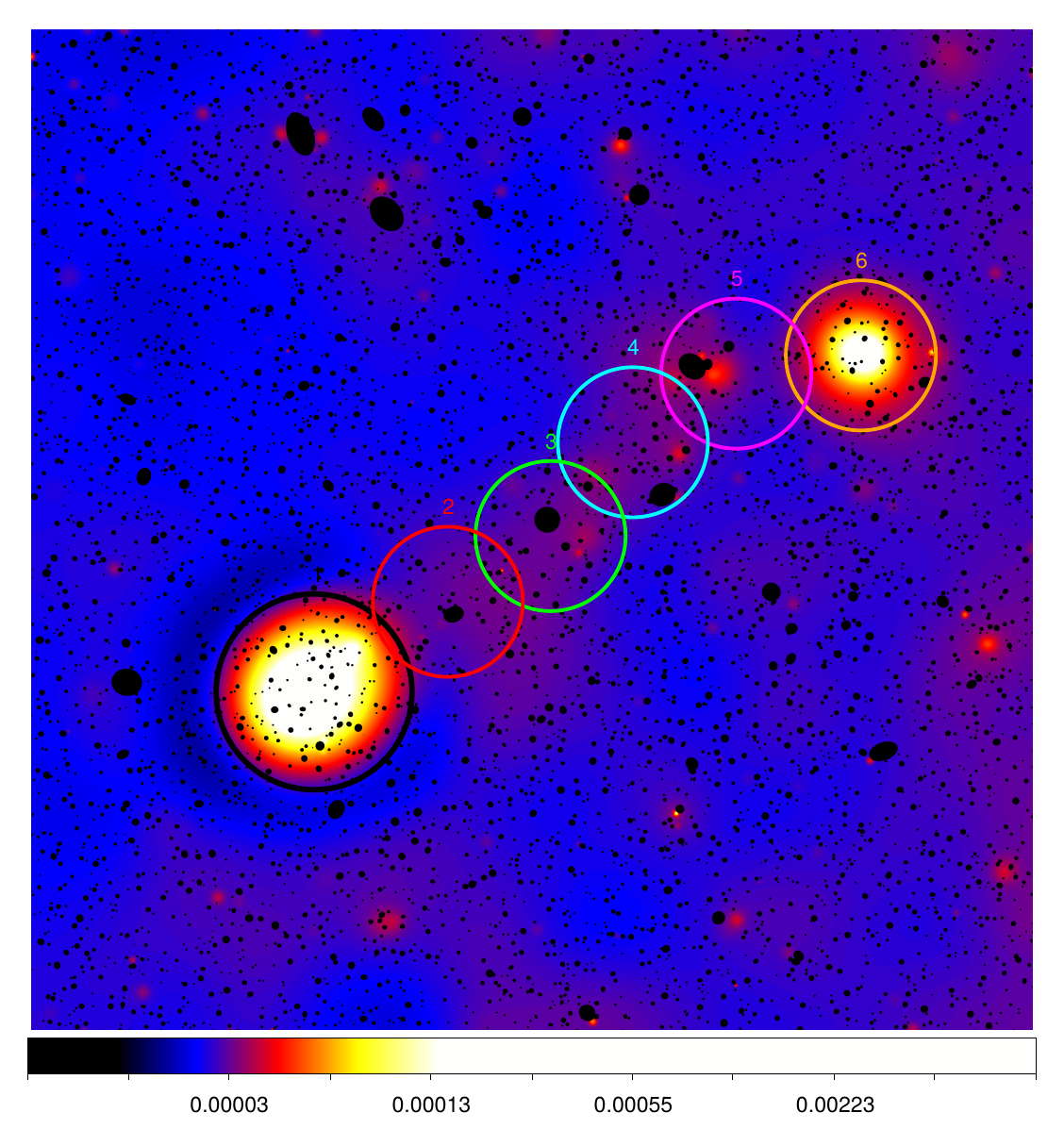}
   \includegraphics[width=\hsize]{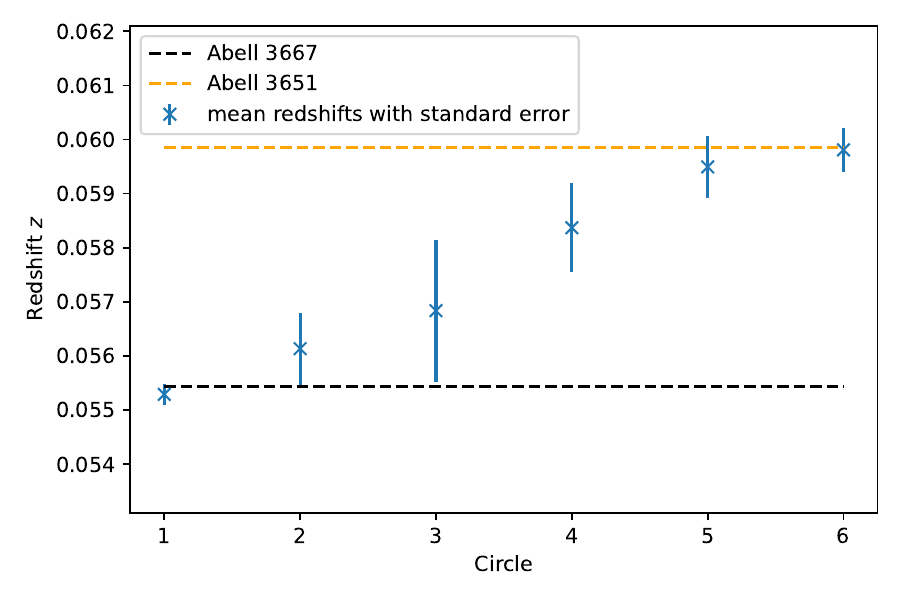}
    \caption{Selection and redshifts of galaxies in the filament region. Top: Data-reduced and wavelet-filtered image with six circles from A3667 to A3651. The galaxies were taken from NED for each circle. The galaxies inside the circles were considered in the redshift analysis. Bottom: Mean redshift of the galaxies, with close-by redshifts for each circle. A progression from the redshift of A3667 to the redshift of A3651 is visible.}
    \label{Fig:redshift_progression}
\end{figure}

For the redshift analysis of the sources between the clusters, galaxies with known redshifts and $z<\num{0.1}$ were extracted from the \textit{NASA/IPAC~Extragalactic~Database~(NED)\footnote{\url{https://ned.ipac.caltech.edu/} The NASA/IPAC Extragalactic Database (NED) is funded by the National Aeronautics and Space Administration and operated by the California Institute of Technologys.}}. The top panel of Figure \ref{Fig:redshift_progression} shows six circles from A3367 to A3651. The circles have a radius of max($\SI{23}{arcmin}$, cluster $R_{200}$). We considered the galaxies inside these circles in the following analysis. Since the circles overlap partially, we ensured that no galaxy was counted twice by always assigning the galaxies in the overlapping regions to the circle farther to the left. The number of the galaxies per circle can be found in \cref{tab:galaxies_number}.

\begin{table}
   \caption{Number of galaxies per circle. The defined circles are shown in \cref{Fig:redshift_progression} top.}
      \label{tab:galaxies_number}
  $$ 
      \begin{array}{ll}
         \hline
         \noalign{\smallskip}
         \textnormal{Circle} &  \textnormal{Number of galaxies} \\
         \noalign{\smallskip}
         \hline
         \noalign{\smallskip}
         1 & 534 \\
         2 & 49 \\
         3 & 15 \\
         4 & 20 \\
         5 & 49 \\
         6 & 70 \\
         \noalign{\smallskip}
         \hline
      \end{array}
  $$ 
\end{table}

The algorithm called density-based spatial clustering of applications with noise \texttt{(DBSCAN)} \citep{DBSCAN} was used to identify several categories of galaxies with similar redshifts. A category was defined via the two parameters $minPts = 3$ and $dist = 0.004$\footnote{\textit{minPts} is the minimum number of data points inside a circle with radius \textit{dist} to define a core point. A category is identified as a chain of points that are reachable inside a distance smaller than \textit{dist} from a core point.}.

\begin{figure}
   \centering
   \includegraphics[width=\hsize]{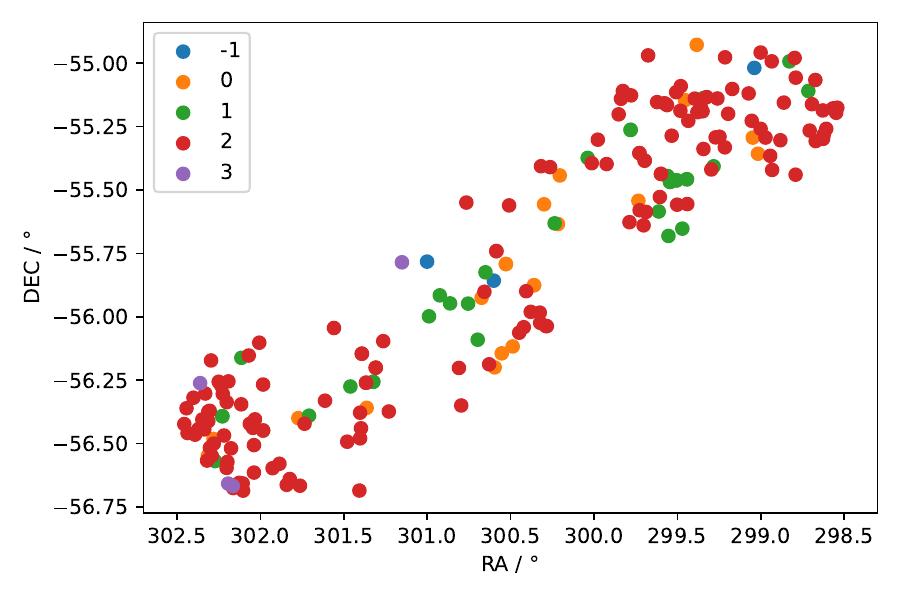}
   \includegraphics[width=\hsize]{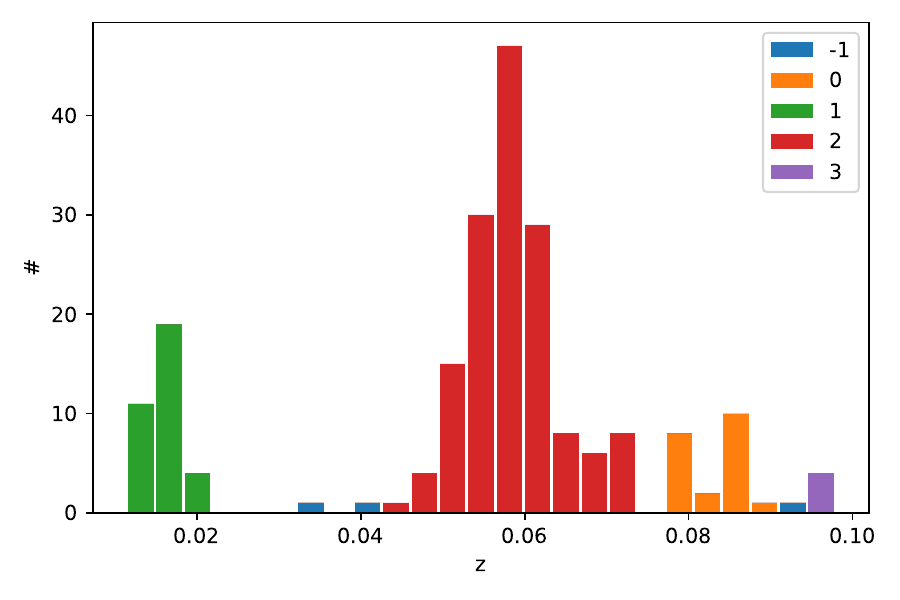}
    \caption{Spatial (top) and redshift (bottom) distribution of galaxies in the region of the galaxy overdensity between A3667 and A3651. The different colors (and corresponding label numbers) show categories of galaxies with a similar redshift identified with the \texttt{DBSCAN} algorithm. By comparing the top and bottom plots, the spatial location of galaxies with certain redshifts can be seen. Galaxies with the label $-1$ are outliers that could not be identified with one of the categories.}
    \label{Fig:DBScan_plot}
\end{figure}

The top panel of Figure \ref{Fig:DBScan_plot} shows the spatial distribution of the galaxies in circles 2-4. A redshift histogram of these galaxies is shown in \cref{Fig:DBScan_plot} bottom. The categories identified by \texttt{DBSCAN} are shown by different colors in the histogram and in the spatial distribution plot. Based on \cref{Fig:DBScan_plot}, it is possible to directly compare where galaxies with certain redshifts are located spatially. Most of the galaxies, namely those plotted in red, feature a redshift similar to that of the galaxy clusters A3667 and A3651. The redshifts of these galaxies are defined from now on as close-by redshifts. A detailed discussion of the plots follows in Sect. \ref{sec:discussion}. The same plot and histogram for all NED galaxies in the FoV can be found in \cref{sec:app_zdistri}.

Furthermore, the redshift distribution inside the individual circles was investigated. Based only on the galaxies with the previously defined close-by redshifts, we calculated the mean redshift in each circle. We plot it with its standard error in the bottom panel of \cref{Fig:redshift_progression}. Starting from cluster A3667, the mean of the redshifts increases to the redshift of the cluster A3651. The different standard errors originate from the different number of galaxies per circle (see~\cref{tab:galaxies_number}).
The MCXC redshifts of the clusters, represented by the horizontal dashed lines, are not exactly equal to the mean values of circles~1 and 6 because a different number of galaxies were used to calculate the mean redshift. While we have 519 close-by galaxies in circle~1 and 55 close-by galaxies in circle~6, the MCXC redshifts were calculated from 162 galaxies for A3667 and 79 galaxies for A3651 \citep{ROSAT_catalog}. Since it is not our purpose to redetermine the cluster redshifts and because the values are consistent with each other, we continue to use the MCXC values as the default in this work.

\subsection{Comparison to the \textit{SLOW} simulation}
\label{sec:simulations}

\begin{figure}[htbp]
    \centering
    \includegraphics[width=0.9\hsize]{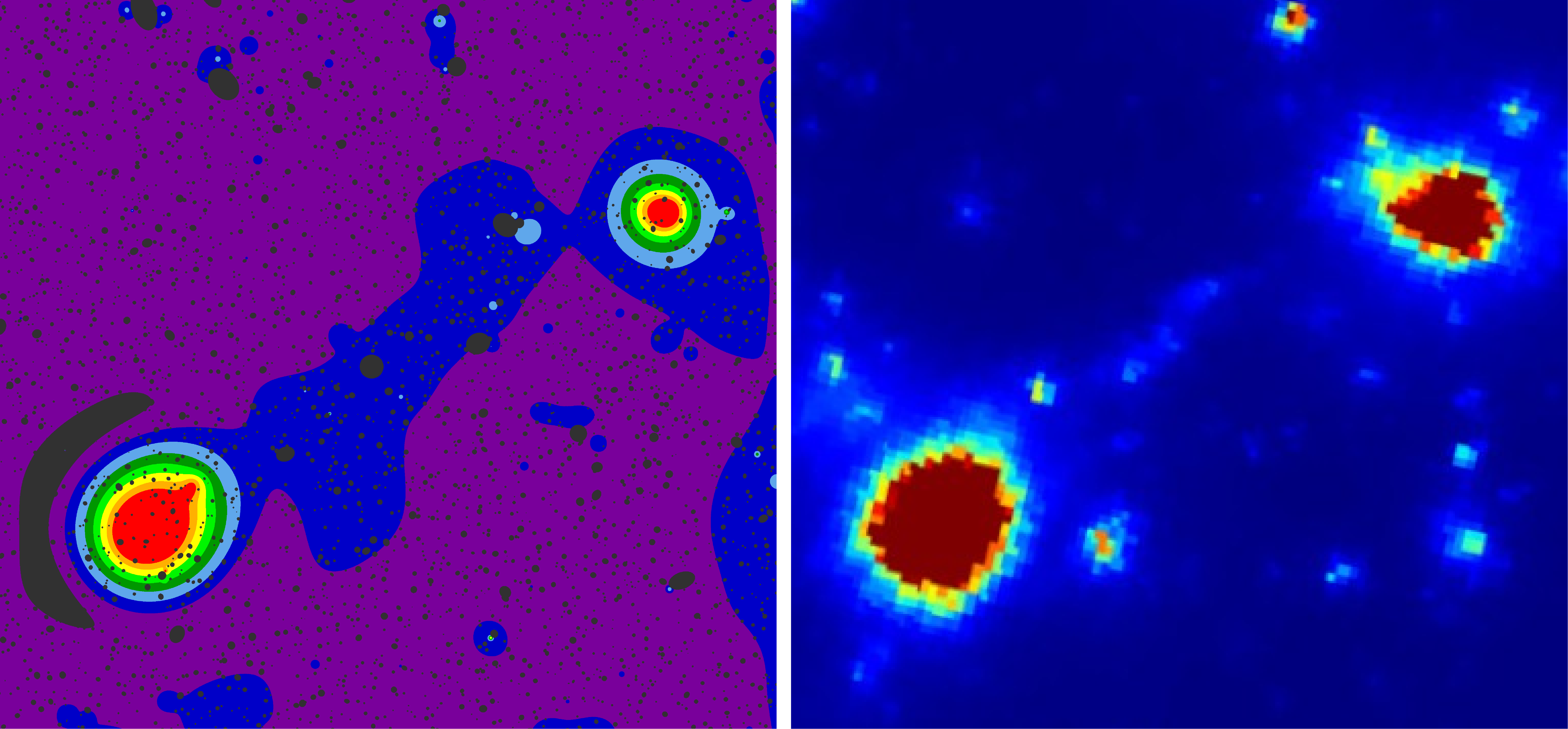}
    \caption{Comparison of data and simulation. Left: eRASS:4 X-ray image. Right: Same image section in the constrained simulation \textit{SLOW} (\citealt{SLOW}; Seidel et al.~in prep.) The two clusters, their ellipticities and extensions toward the filament region, and the filament itself are reprocuded in the mock observation.}
    \label{Fig:sim_compare}
\end{figure}

Figure \ref{Fig:sim_compare} again shows the eRASS:4 X-ray image on the left and the same image section from the constrained hydrodynamical simulation \textit{Simulating the LOcal Web (SLOW)} \citep{SLOW} on the right. The two clusters A3667 and A3651 are clearly visible in the \textit{SLOW} image, also with their ellipticities and extensions toward the filament region, and a thin filament spine can be seen between the clusters. While the simulation was constrained to reproduce the main features of the Local Universe, it is striking that the filament is reproduced as well. \change{However, although the clusters are connected by a filament and move toward each other in comoving space, they are not gravitationally bound to each other in this simulation (Seidel et al.~in prep).} Using cylindrical shells around the filament spine in the mock observation, we extracted the electron density profile and estimated a mean temperature of $T = \SI{0.53}{keV}$ in the cylinder. \change{The comparison to the results obtained with a spectral analysis of the eROSITA data is described in Sect. \ref{sec:discussion}}.

\subsection{Filament length}
\label{sec:filament_length}
\begin{figure}
   \centering
   \includegraphics[width=\hsize]{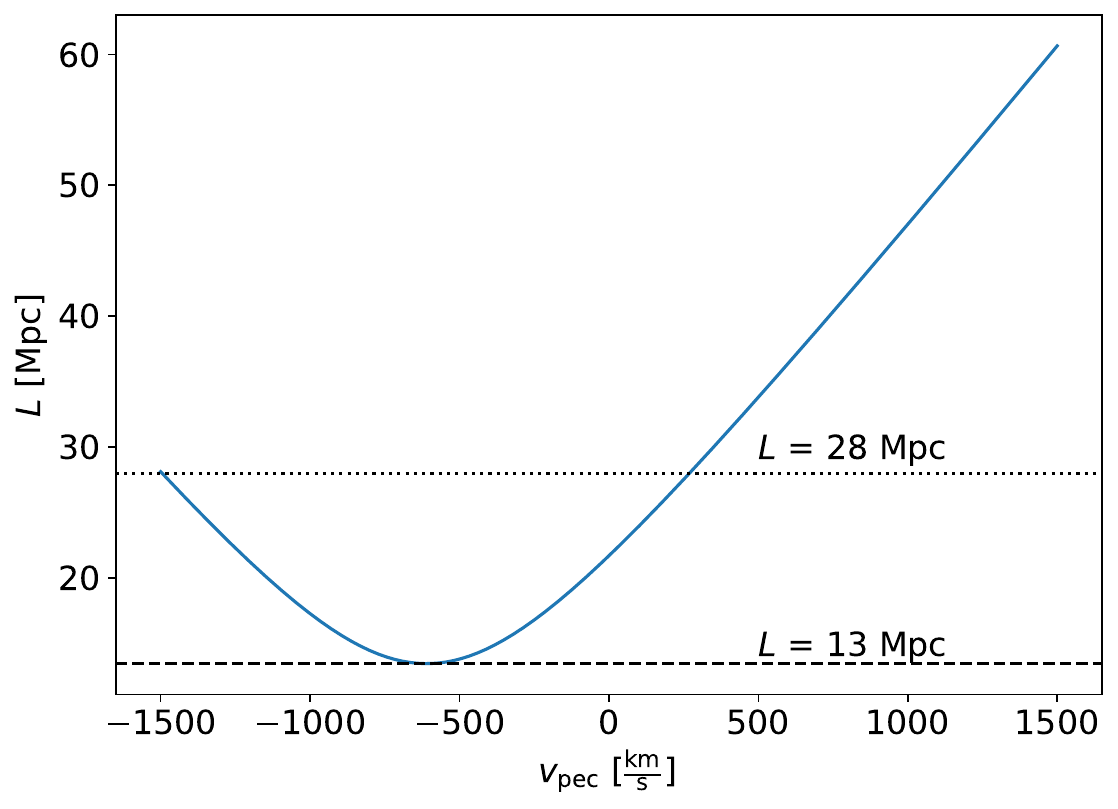}
   \includegraphics[width=\hsize]{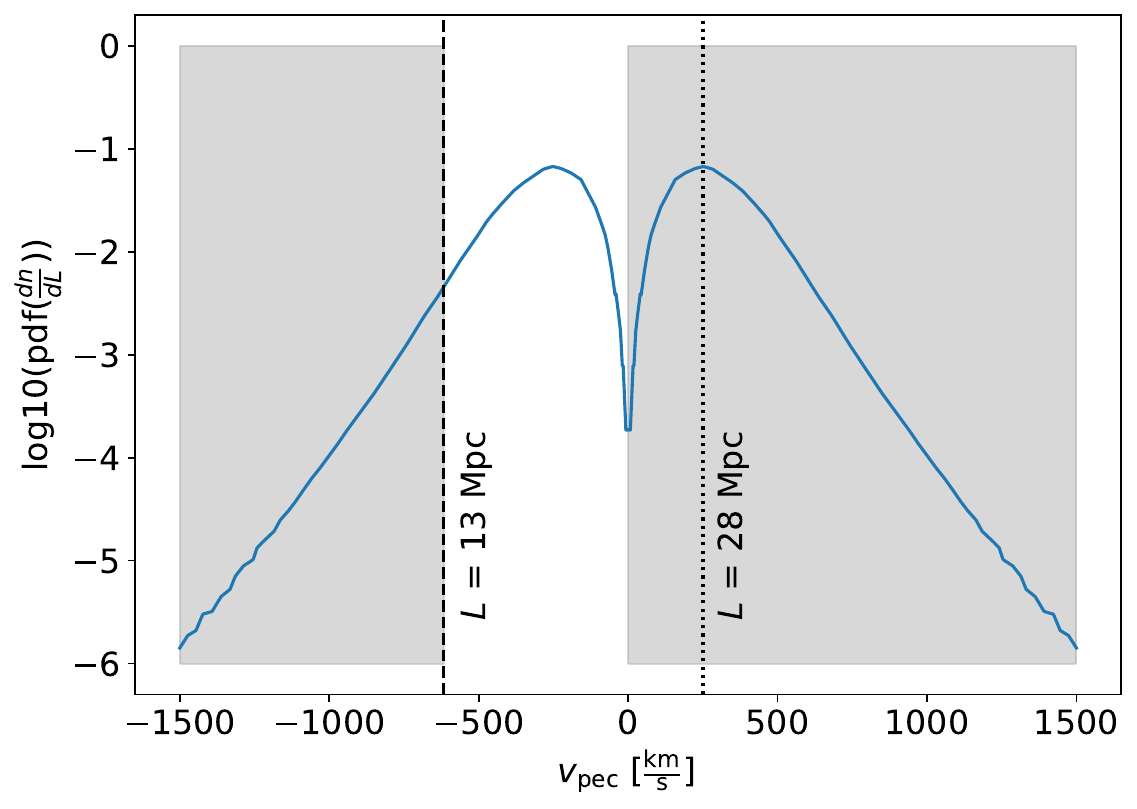}
    \caption{Impact of peculiar velocity on the filament length. Top: Filament length $L$ in dependence of the line-of-sight peculiar velocity $v_{\mathrm{pec}}$ of one cluster.  Bottom: Probability density function for $L$ as a function of $v_{\mathrm{pec}}$. The shaded gray regions show velocity ranges in which the clusters move toward each other. The minimum and most probable lengths are highlighted.}
    \label{Fig:filament_length}
\end{figure}

The angular separation between the cluster centers of A3667 and A3561 is $\SI{3.34}{\degree}$, which corresponds to a transverse filament length of ${\sim}\SI{13}{\mega pc}$, using the redshift of A3667.  This lower limit for the filament extent disregards the different redshifts of the two clusters, and therefore, the line-of-sight distance. The challenge of computing the three-dimensional distance between the two clusters stems from the absence of data regarding their peculiar velocities. The top panel of Figure \ref{Fig:filament_length} shows the length $L$ (three-dimensional proper distance from one cluster center to the other) of the filament in dependence of the line-of-sight peculiar velocity $v_{\mathrm{pec}}$ of one cluster. The second cluster is presumed to exhibit an identical peculiar velocity, but with the opposite sign. $v_{\mathrm{pec}} = 0$ means that the observed redshifts of the clusters are equal to their cosmological redshifts, which gives a length of $\SI{22}{\mega pc}$. A positive $v_{\mathrm{pec}}$ suggests that A3667 is receding from the observer while A3651 approaches, leading to a difference between the cosmological redshifts greater than that of the observed redshifts, and consequently, to an increased filament length.  Conversely, the reverse is true for a negative $v_{\mathrm{pec}}$, but only up to the minimum of the graph. This minimum corresponds to a velocity where the cosmological redshifts of the clusters match, yielding a transverse length of $\SI{13}{\mega pc}$. Beyond this point in the direction of more negative $v_{\mathrm{pec}}$, A3667 becomes the more distant cluster and A3561 the closer one, thereby inverting the geometrical arrangement and once again elongating the filament.

By employing the probability density function (pdf) from \citep{Boulliot_2015} for pairwise cluster velocities with separations $< 15 h^{-1}\,\si{\mega pc}$, we created the graph in the bottom panel of \cref{Fig:filament_length}. This graph presents the pdf for the filament length as a function of $v_{\mathrm{pec}}$. The shaded gray regions show velocity ranges in which the cluster geometry indicates mutual convergence, which is plausible for clusters that are connected by a filament. When we incorporate this prior, the likelihood peaks at a length of $\SI{28}{\mega pc}$, with the 16th and 84th percentiles bracketing $\SI{25}{\mega pc}$ and $\SI{32}{\mega pc}$, respectively. The line-of-sight inclination angle, derived from this length, is $\SI{61}{\degree}$ (with the 16th and 84th percentiles at $\SI{58}{\degree}$ and $\SI{65}{\degree}$, respectively).

\subsection{Spectral analysis}
\label{sec:spectral_analysis}
\change{In order to characterize the gas temperature and metallicity of the filament, we performed a spectral analysis following the procedure described in \citet{3391_95}. Only the main steps of the procedure are repeated here.}

\change{Using the \texttt{eSASS} task \texttt{srctool}, spectra, we extracted ancillary response files (ARFs) and response matrix files (RMFs) for the source and background regions. The location of these regions are shown in \cref{fig:transverse_SB} (upper panel). All seven TMs were used, the TMs with on-chip filters in the energy range $\num{0.3}$--$\SI{9.0}{\kilo\electronvolt}$, and the other two TMs were restricted to $\num{0.8}$--$\SI{9.0}{\kilo\electronvolt}$. Using \texttt{XSPEC} \citep{xspec}, we fit a model including the instrumental background (PIB), components for the unabsorbed local hot bubble (LHB), absorbed Milky Way halo (MWH), absorbed unresolved cosmic X-ray background (CXB), and absorbed thermal source (filament). The total hydrogen column density $N_{\text{H,tot}}$ from \citet{Willingale_13} from the center of each extraction box was used for the corresponding absorption component.}

\begin{figure}[htbp]
   \centering
   \includegraphics[width=\hsize]{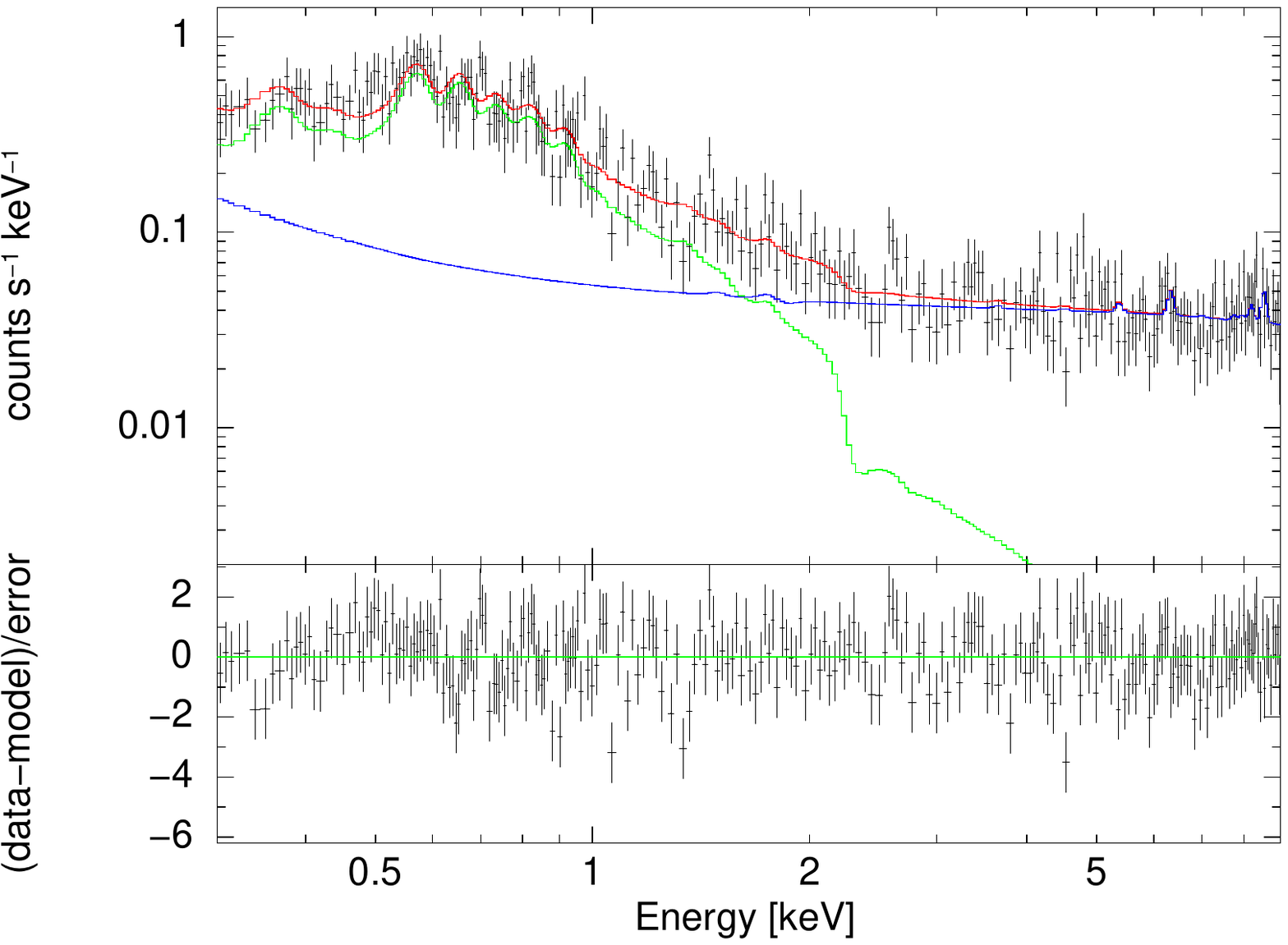}
    \caption{\change{eROSITA spectrum and fitted models for TM1. The spectral data are plotted in black, and the solid lines represent the fitted model. Blue represents the instrumental background, green the sky component (filament + sky background), and red the total model.}}
    \label{Fig:spectrum_example}
\end{figure}

\change{First, we fit the normalizations of the background region, with a fixed temperature and metallicity for the LHB and MWH, and a fixed power-law slope for the CXB. The values for the fixed parameters can be found in Table~3 in \citet{3391_95}. Second, the source region was fit, for which we again thawed the background normalizations, but used the previously constrained normalizations as starting values. The remaining background parameters were kept fixed. An example for the extracted spectrum and fitted model for one TM is shown in \cref{Fig:spectrum_example}.}

\change{Since the results of the fit in a low surface brightness region can be very sensitive to the modeling choices, we performed multiple fits in which we varied the $N_{\text{H}}$, the energy range for the fit, and the redshift of the filament. Furthermore, we performed two fits using two different background regions (background~north and background~south, as shown in the top panel of \cref{fig:transverse_SB}). The results are very stable against the changes in the parameters, but show small differences for the two background definitions. A description of the tests and the results for all fitting parameters can be found in \cref{sec:app_spectral}.}

\change{As final values, we report the results of the fitting procedure using background~south, but we increased the uncertainties such that they include the results of the fitting procedure using background~north. We chose background~south because the X-ray emission in this region box is more representative for the whole background and is more constant than the emission in background~north (see the values of the E~boxes and A~boxes in \cref{Fig:SB_boxes} bottom). It also shows much less variation in $N_{\text{H}}$. Nevertheless, we cannot exclude other systematic differences between the different background regions and present the most conservative estimate with the combined error bars of both modeling choices. The final results are $T=(\num{0.91}_{-0.11}^{+0.07})\,\mathrm{keV}$ and $Z=(\num{0.10}_{-0.08}^{+0.05})\,\mathrm{Z_{\odot}}$.}

\subsection{Integrated gas properties and electron density}
\label{sec:filament_gas}
\begin{figure}[htbp]
        \centering
    \includegraphics[width=0.98\hsize]{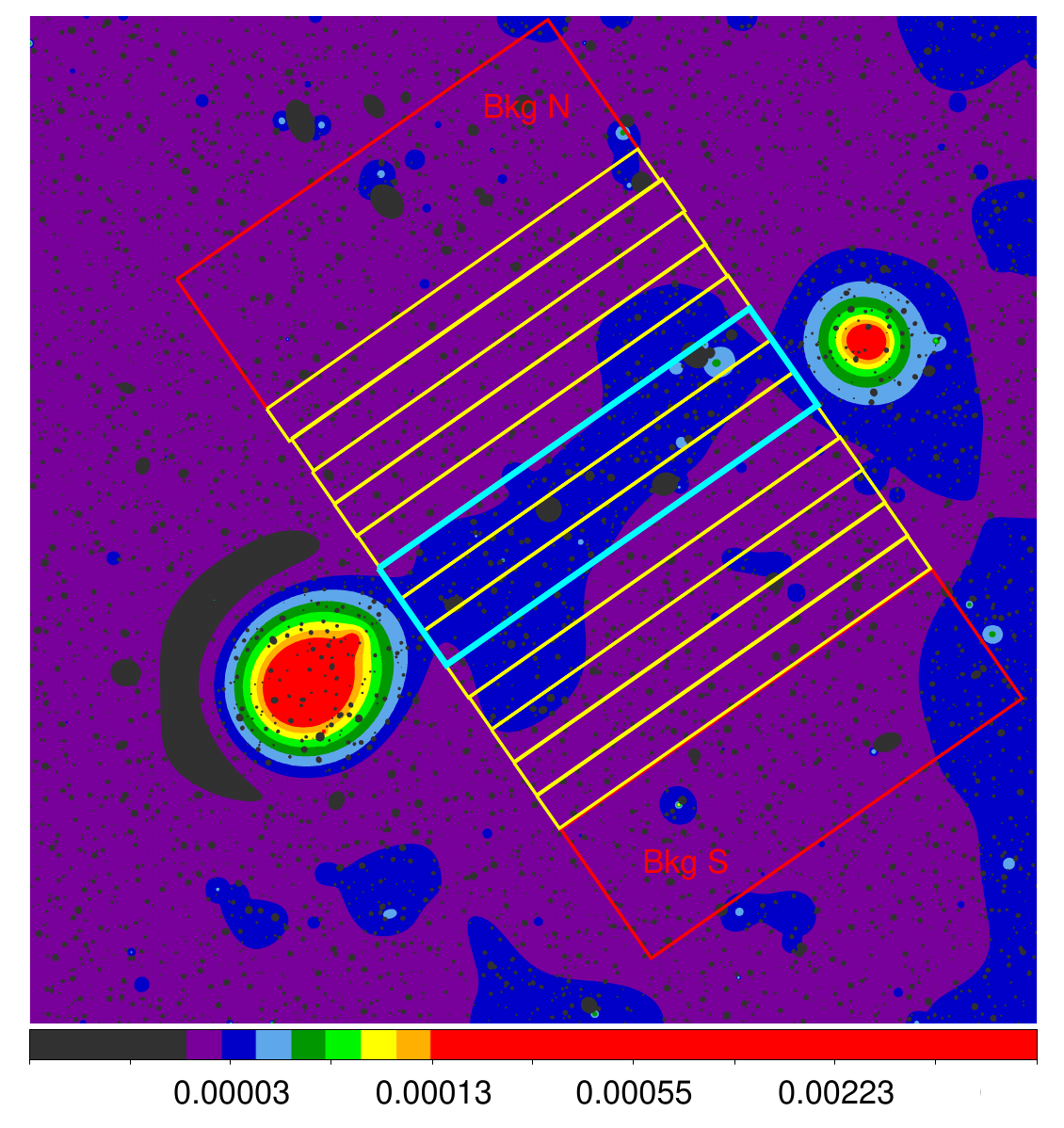}
        \includegraphics[width=\hsize]{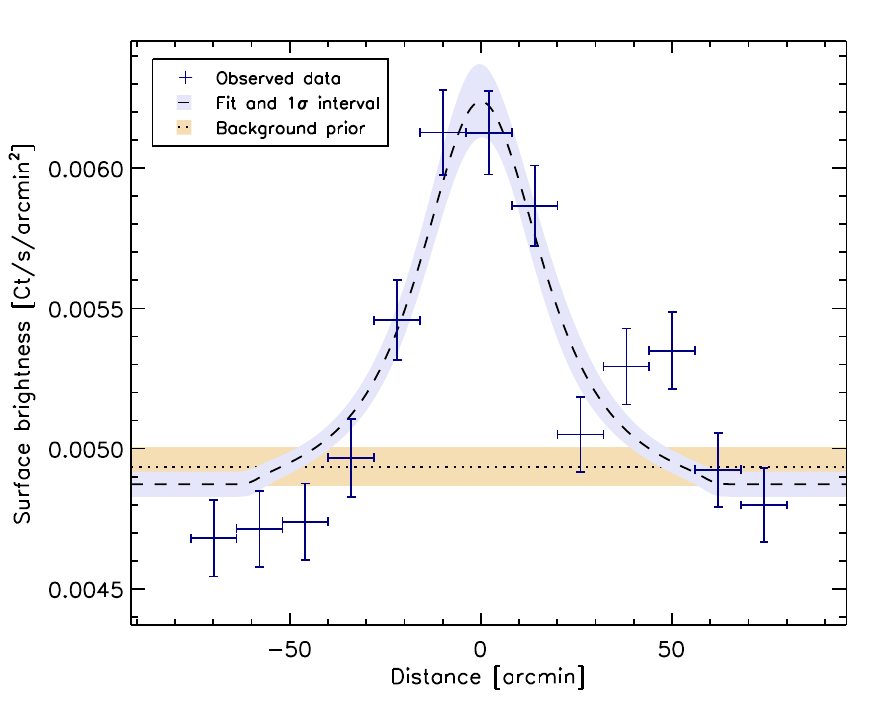}
        \caption{Transverse SB profile of the filament. Top: Configuration of the boxes used to extract a transverse SB profile (yellow). The length of the boxes, $\SI{2.3}{deg}$, was set to minimize the contribution from cluster outskirts while retaining most of the filament contribution. The box width is $\SI{12}{arcmin}$. The SB background is constrained from both red boxes. \change{The same background boxes are used for the spectral analysis. The cyan box is the combination of the three central yellow boxes and was used as source extraction region for the spectral analysis}. Bottom: Resulting SB data points with a truncated $\beta$-model fit with $\beta=2/3$ and $R_\mathrm{max}=\SI{60}{\arcmin}$. Positive distances are toward Bkg S, and negative distances are towards Bkg N.}
        \label{fig:transverse_SB}
\end{figure}

In order to estimate \change{further} gas properties of the filament, we carried out an SB analysis similar to that of Sect. \ref{sec:filament_SB}, but with a different box placement to create a transverse SB profile of the filament. A truncated 2D $\beta$-model (with a maximum radius of $\SI{60}{\arcmin}$ and a fixed $\beta$=2/3) was projected along the line of sight and fit to the observed data points. In doing so, we assumed an angle between the filament and the plane of the sky of $\SI{61}{\degree}$ based on the results of Sect. \ref{sec:filament_length}. We opted here for a truncated model since integrating to infinity the surface density of such a $\beta$-model would result in a diverging total gas mass. The configuration of the boxes and the resulting fit is shown in \cref{fig:transverse_SB}.

\change{Using the filament temperature of $T_\mathrm{X}=\SI{0.91}{keV}$ and metal abundance of $Z=\SI{0.1}{Z_\odot}$ as obtained from the spectral analysis}, we converted the count-rate emissivity profile into flux, electron, and gas mass density profiles. Integrating over the profiles, we estimated a total filament count-rate of $\SI{9.2\pm1.0}{ct\,s^{-1}}$ and a flux of \change{$F_\mathrm{X}=\SI{7.4(8)e-12}{erg\,s^{-1} cm^{-2}}$} within the $\SI{2.57}{deg}$ separation of the cluster $R_\mathrm{200}$ radii. This translates into a gas mass of \change{$M_\mathrm{g}=\SI{2.7(0.2)e+14}{M_\odot}$} and a central baryon overdensity in the filament of \change{$\delta_0 := \frac{\rho_{\text{gas}}-\overline{\rho}_{\text{gas}}}{\overline{\rho}_{\text{gas}}} = \num{215}^{+23}_{-21}$, where $\overline{\rho}_{\text{gas}}$ is the average baryon density of the Universe}. Nevertheless, some of these quantities strongly depend on our assumptions about the gas emissivity and the inclination of the filament, and we also estimated more conservative errors by varying some of our assumptions: the inclination of the filament in the range $\SIrange{58}{65}{\degree}$ (as estimated in Sect. \ref{sec:filament_length}), \change{the gas temperature between $\num{0.80}$ and $\SI{0.98}{keV}$, the gas metallicity between $\num{0.02}$ and $\SI{0.15}{Z_{\odot}}$ (as estimated in Sect. \ref{sec:spectral_analysis})}, and the truncation of the $\beta$-model between $R_\mathrm{max}=51$ and $\SI{69}{{}^\prime}$ (i.e., 15\% of our baseline value). This resulted in final estimates of \change{$F_\mathrm{X}=\num{7.4(12)e-12}\,\mathrm{erg\,s^{-1}\,cm^{-2}}$, $M_\mathrm{g}=(2.7^{+1.4}_{-0.8})\times 10^{14}\,\mathrm{M_\odot}$} and \change{$\delta_0=215^{+86}_{-50}$}.\change{The uncertainties here were determined by varying the assumptions in the ranges given above and taking the parameter combinations that vary most from the central values.}

\section{Discussion and conclusion}
\label{sec:discussion}
On the basis of the analysis carried out in Sect. \ref{sec:analysis}, we can now evaluate to which extent the visible structure between the galaxy clusters is indeed a filament.

The visual inspection of the eRASS:4 X-ray image (e.g., \cref{Fig:SystemOverview}) shows a very clear excess of emission in between the two clusters. Starting northwest of A3667, the apparent structure crosses the image diagonally east of A3651. No other structures of similar shape and size are visible in the image.

To obtain quantitative results, we analyzed the surface brightness of the filament region and the surrounding regions (see \cref{Fig:SB_boxes}). Four background regions parallel to the filament, each split into three boxes, vary by about $\SI{20}{\percent}$. Because no systematic trend is visible, and when we take into account that the error bars overlap partially, this variation can be considered as a statistical fluctuation. However, the three boxes within the filament are clearly above the background values. Therefore, the boxes within the filament were combined to one SB value, and the same was done for the background boxes. When we compare these two values, the emission inside the filament is $\SI{30(3)}{\percent}$ higher than in the background at a significance of $\SI{11}{\sigma}$. Since the relative difference between filament and background is larger than the relative variation within the background or filament boxes and because the significance level is exceptionally high, this analysis clearly indicates that it is  extremely unlikely that the filament structure is just a random variation in the data.

Up to this point, it can be said that the region between the two clusters clearly is a structure with unusually high emission. In particular, regions lying outside the $3R_{200}$ of galaxy clusters do not normally show this behavior, and only faint emission is often seen between $R_{200}$ and $3R_{200}$ as well because the X-ray surface brightness drops below various fore- and background components at large radii \citep{OutskirtsPaper}.

Particular attention should still be paid to Abell~S0840, however, which according to the MCXC is a galaxy group between A3667 and A3651 and might cause the enhanced emission. However, the SB~profile (see \cref{Fig:SB_S0840}) of the reduced \textit{XMM-Newton} data shows that this is not the case. It becomes clear that significant X-ray emission can only be detected up to a radius of $\SI{4}{arcmin}$ at most. This is much smaller than the previously estimated $R_{500} = \SI{15.6}{arcmin}$, at which no enhanced emission is visible at all. Together with the analysis in \cref{sec:app_S0840}, this quantitatively shows that the flux of Abell~S0840 beyond these $\SI{4}{arcmin}$ is negligible, and it justifies the decision to just excise a $\SI{4}{arcmin}$ sized region for the SB analysis of the filament.

Moreover, cosmic-web filaments can also be traced by the trail of galaxies. Thus, a 2MASS~map (\cref{Fig:2MASS_both} left) was used to study the spatial distribution of galaxies in this region of the sky. The 2MASS~map indeed not only shows a significant galaxy overdensity in the cluster regions, as expected, but also in the region between the two clusters. An overdensity of galaxies is observed along the entire filament. In particular, a galaxy overdensity is evident east of A3651. At this position, a number of sources were excised in the course of the point-source removal, including one extended source east of A3651. This is the radio and X-ray source PKS 1954-55 with a redshift of $z = \num{0.05845}$ \citep{PKS}. This might be another galaxy group at a similar redshift as A3651 and part of the filament structure.

However, it is unclear whether both the other galaxies in the 2MASS~map and the observed filament are spatially located between the galaxy clusters or if they only appear in the projection from the sources to the observer. Therefore, an analysis of the redshifts was carried out.

Galaxies from NED were studied in six circular regions from A3667 to A3651. The \texttt{DBSCAN} algorithm was used to group the galaxies into five redshift categories. The colors in the top and bottom panels of \cref{Fig:DBScan_plot} match, so that the two plots allowed us to directly compare where galaxies with certain redshifts are located spatially. In the top panel of \cref{Fig:DBScan_plot}, the regions of the galaxy clusters themselves are not shown but only the region between them to avoid overcrowding the plot with ${\sim} \num{700}$ more data points. The
same plot and histogram for all NED galaxies in the FoV can be found in \cref{sec:app_zdistri}. The redshifts of most of the galaxies are similar to those of clusters A3667 and A3561 (red). They are distributed throughout the entire filament structure, with higher densities towards the clusters. There are a few outliers (blue) and galaxies with higher redshifts (orange and purple), but these galaxies are not restricted to a specific spatial location. The galaxies with redshifts similar to the redshift of S0840 (green) indeed show a small spatial accumulation, but it is very small compared to the red galaxies near the other two clusters, and green galaxies can also be found throughout the entire filament structure and are not only restricted to the location of S0840. Hence, the presence of a galaxy cluster in the middle of the filament is not clear from this analysis either.

Furthermore, the arithmetic mean of the redshifts of the red galaxies was determined for each of the four regions. \cref{Fig:redshift_progression} indeed shows that there is a progression of redshifts. Starting from the redshift of A3667, which is the lowest redshift, the redshift increases along the filament to the redshift of A3651, which is the highest redshift.

The fact that the galaxies, especially those whose redshifts are similar to the redshift of the clusters, show a higher density toward the clusters might indicate the flow of material toward the cluster \citep{Bond}. This theory is also supported by the northwestern substructure of A3667, which is aligned in the direction of the filament \citep{Gasperin_A3667}.
Although not all galaxies have the same redshift as the clusters and the 2MASS map may show projection effects, the trail of galaxies between the clusters therefore makes it plausible that the visible X-ray emission between the clusters originates from a redshift similar to the cluster redshift.
In addition, the main features of the galaxy clusters and the enhanced emission between them are reproduced by the constrained simulation \textit{SLOW}, as shwon in \cref{Fig:sim_compare},

Combining all these pieces of evidence, we conclude that we have discovered an X-ray filament connecting the two galaxy clusters. We next estimate its gas properties and compare the results with simulations.

\change{The spectral analysis of the filament region results in values for the temperature and metallicity of $T=(\num{0.91}_{-0.11}^{+0.07})\,\mathrm{keV}$ and $Z=(\num{0.10}_{-0.08}^{+0.05})\,\mathrm{Z_{\odot}}$.  According to \citet{Migkas_eROSITA_T}, for this $T \sim \SI{1}{keV}$, no offset between eROSITA and XMM/Chandra is expected. Our $T$ estimate is therefore robust and insensitive to the instrument that is used. Our findings are consistent with previous eROSITA filament observations (e.g.,\ \citealt{3391_95} for the Abell 3391/95 system and \citealt{Tanimura+2022} for a stacking analysis). However, these temperatures are higher than typically expected for pure WHIM filaments (e.g.,\ \citealt{TNG_Daniela}). Nevertheless, these high values occur in the large scatter found in numerical simulations.}

\begin{figure}[htbp]
    \centering
    \includegraphics[width=\hsize]{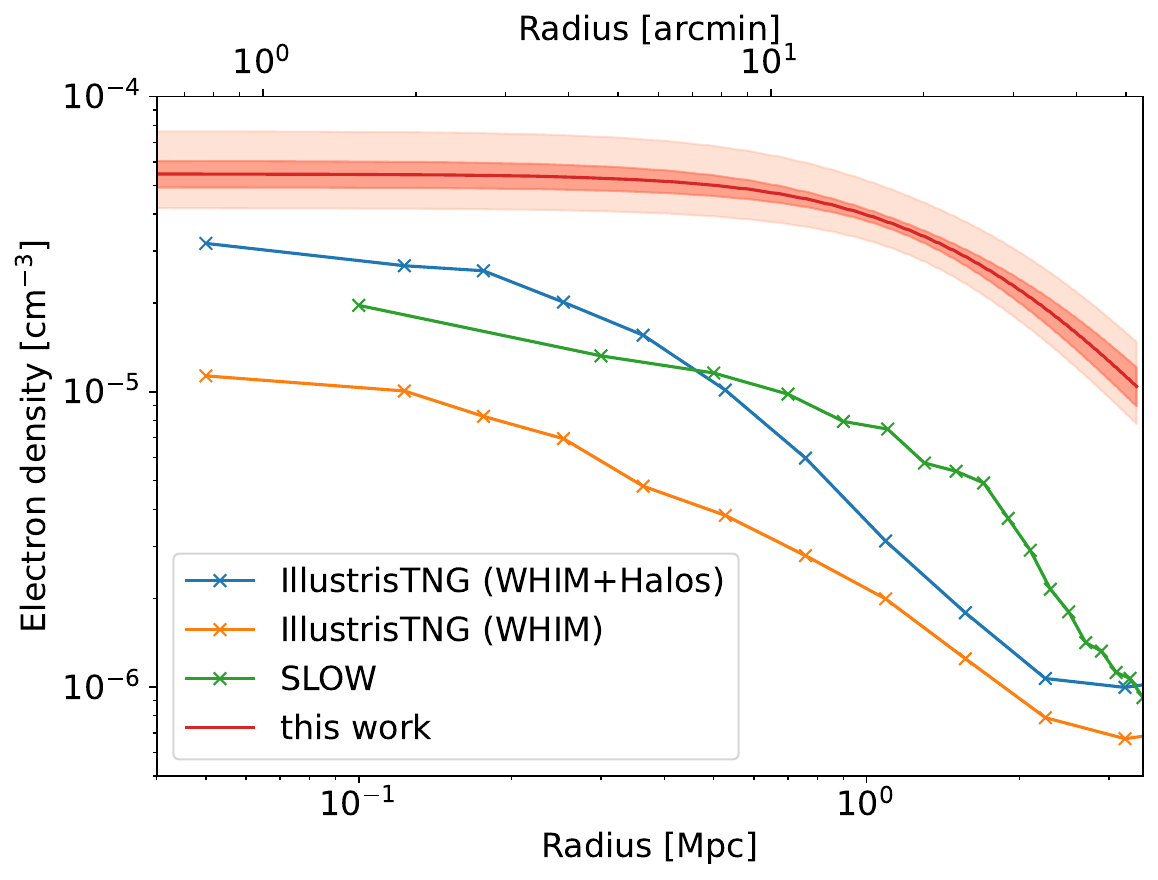}
    \caption{Comparison of the electron density profiles of this work, the reproduced filament in the SLOW mock observation, and the $L>\SI{20}{Mpc}$ WHIM filaments at $z=0$ \citep{TNG_Daniela} in the IllustrisTNG simulation. The uncertainty intervals of this work divide into the (smaller) statistical uncertainty of the data and the (larger) systematic uncertainty \change{including the temperature and metallicity ranges obtained from the spectral analysis and varying the assumptions of the inclination angle and truncation radius of the $\beta$-model}. However, these uncertainties are not independent of each other.}
    \label{Fig:ne_profiles}
\end{figure}

We fit a truncated $\beta$-model to the filament emission and \change{used the temperature and metallicity from spectral analysis} to estimate a total X-ray flux of \change{$F_\mathrm{X}=\num{7.4(12)e-12}\,\mathrm{erg\,s^{-1}\,cm^{-2}}$}, which is accompanied by a gas mass of \change{$M_\mathrm{g} = (2.7^{+1.4}_{-0.8})\times 10^{14}\,\mathrm{M_\odot}$} and a central baryon overdensity of \change{$\delta_0 = 215^{+86}_{-50}$}.
This analysis provides further insights into the filament properties, emphasizing the typical gas mass while addressing the central overdensity: The recorded value of ${\sim}\num{200}$ appears notably elevated, as it typically corresponds to the overdensity of virialized halos. This might come from the wide error range in our analysis. However, it is plausible that this discrepancy is partly due to the blending of genuine filament emission with emission from collapsed halos that remain unresolved, \change{including possible contributions from small groups and clumps residing in the filament.}
Furthermore, we compared the electron density profiles of the eROSITA observation, of the SLOW mock observation, and of average filaments with $L>\SI{20}{Mpc}$ at $z=0$ \citep{TNG_Daniela} in the IllustrisTNG simulation \footnote{\href{https://www.tng-project.org/}{https://www.tng-project.org/}} in \cref{Fig:ne_profiles}. From the latter, we showed two profiles: One profile with pure WHIM, and another profile that included halos, which in this case are smaller-scale collapsed structures, such as galactic halos. Since we did not remove any emission from possible subhalos in the filament in the observation, we compared this to the IllustrisTNG curve with halos. While the profile of average filaments in the simulation shows a still lower electron density, the actual scatter is larger, and individual filaments can have higher values of the electron density. With the given case that it is possible to observe the filament in X-ray emission, it is plausible that it will be at the upper end of the scope. Moreover, the IllustrisTNG curve drops faster, while the observed profile remains constant up to a larger radius. This is because we did not follow the exact spine of the filament in our analysis, but only analyzed the emission within a straight line between the two clusters. When the spine is not precisely defined, this has the effect of smearing the profile.
The profile of the SLOW filament lies below the IllustrisTNG curve for small radii, but also shows the smearing effect for larger radii. Its electron density is considerably lower than is observed, but it also shows a lower emissivity when compared in \cref{Fig:sim_compare}.

In conclusion, through a surface brightness analysis of the reduced eRASS:4 data from the eROSITA X-ray telescope, a filament structure was discovered in the energy range $\num{0.3}$--$\SI{2.0}{\kilo\electronvolt}$ from the galaxy cluster A3667 to the galaxy cluster A3651. An additional \textit{XMM-Newton} analysis excluded the possibility that the emission originates from the foreground object Abell~S0840. A comparison of the studied region with a 2MASS map and analyses of the redshifts of the galaxies that are inside the filament supported this discovery. Furthermore, the redshift analysis revealed that the galaxies are aligned spatially and in redshift space along the filament. The three-dimensional length of the filament was estimated to be in the range of $\SI{25}{\mega pc} - \SI{32}{\mega pc}$ using the probability density function for pairwise cluster velocities from \citep{Boulliot_2015} and assuming that the clusters move toward each other.
We estimated the key properties of the filament, namely temperature, metallicity, total flux, gas mass, and central baryon overdensity.
The electron density profile was compared to predictions from hydrodynamical simulations, namely to average IllustrisTNG filaments \cite{TNG_Daniela} and to the reproduced filament in the SLOW simulation (\citealt{SLOW}; Seidel et al.~in prep.). While the observed profile is higher than the simulated profiles and has a flatter slope, there are still reasons for the plausibility of our results. With a better-resolved filament spine and comparisons with filaments between high-mass clusters and filaments leading to detectable X-ray emission alone, the electron density profiles are expected to agree better.

Since the observed emission is also outside the $3R_{200}$ of both galaxy clusters, where the cluster outskirts end by definition, it could be the emission from WHIM. One of the scientific goals of the eROSITA mission, to detect hot intergalactic gas in galaxy clusters and in their outskirts and surroundings, would then have been achieved in this example.
In comparison to other well-known galaxy cluster systems with X-ray emission filaments \change{(see the references given in Sect. \ref{sec:introduction})}, this finding is exceptional in terms of its length and significant emission outside the cluster outskirts. Unlike previous discoveries of similar structures limited to a few megaparsec in length or within the outskirts of galaxy clusters, this discovery is, to our knowledge, the first individual filament between two galaxy clusters with a considerable separation and clearly outside the intracluster medium.
\change{Further observations with better spatial resolution, for instance, by \textit{XMM-Newton}, are required to resolve the substructures and clumps within the filament. Then, their contributions can be quantified, and by subtracting them, the filament gas properties can be determined to higher precision.}
The LSS filaments in this energy range that connect galaxy clusters are part of the high-temperature fraction of the WHIM and might contain a significant amount of baryonic matter, thus contributing to the still-unsolved missing baryon problem. A solution to this problem would help us to gain a deeper understanding of the LSS of the Universe.

\begin{acknowledgements}
AV acknowledges funding by the Deutsche Forschungsgemeinschaft (DFG, German Research Foundation) -- 450861021.
CS, AV, and TR acknowledge support from the German Federal Ministry of Economics and Technology (BMWi) provided through the German Space Agency (DLR) under project 50 OR 2112.

KD acknowledges support by the COMPLEX project from the European Research Council (ERC) under the European Union’s Horizon 2020 research and innovation program grant agreement ERC-2019-AdG 882679.

BS acknowledge supported by the grant agreements ANR-21-CE31-0019 / 490702358 from the French Agence Nationale de la Recherche / DFG for the LOCALIZATION project.
\\
This work is based on data from eROSITA, the soft X-ray instrument aboard SRG, a joint Russian-German science mission supported by the Russian Space Agency (Roskosmos), in the interests of the Russian Academy of Sciences represented by its Space Research Institute (IKI), and the Deutsches Zentrum für Luft- und Raumfahrt (DLR). The SRG spacecraft was built by Lavochkin Association (NPOL) and its subcontractors, and is operated by NPOL with support from the Max Planck Institute for Extraterrestrial Physics (MPE).

The development and construction of the eROSITA X-ray instrument was led by MPE, with contributions from the Dr. Karl Remeis Observatory Bamberg \& ECAP (FAU Erlangen-Nuernberg), the University of Hamburg Observatory, the Leibniz Institute for Astrophysics Potsdam (AIP), and the Institute for Astronomy and Astrophysics of the University of Tübingen, with the support of DLR and the Max Planck Society. The Argelander Institute for Astronomy of the University of Bonn and the Ludwig Maximilians Universität Munich also participated in the science preparation for eROSITA.

The eROSITA data shown here were processed using the eSASS software system developed by the German eROSITA consortium.
\\
This publication makes use of data products from the Two Micron All Sky Survey, which is a joint project of the University of Massachusetts and the Infrared Processing and Analysis Center/California Institute of Technology, funded by the National Aeronautics and Space Administration and the National Science Foundation.

This research has made use of the NASA/IPAC Extragalactic Database (NED) which is operated by the Jet Propulsion Laboratory, California Institute of Technology, under contract with the National Aeronautics and Space Administration.

Partly based on observations obtained with \textit{XMM-Newton}, an ESA science mission with instruments and contributions directly funded by ESA Member States and NASA.

This scientific work makes use of the Murchison Radio-astronomy Observatory, operated by CSIRO. We acknowledge the Wajarri Yamatji people as the traditional owners of the Observatory site. Support for the operation of the MWA is provided by the Australian Government (NCRIS), under a contract to Curtin University administered by Astronomy Australia Limited. We acknowledge the Pawsey Supercomputing Centre which is supported by the Western Australian and Australian Governments.

\change{Finally, we thank the anonymous referee for their constructive suggestions that helped us to improve the manuscript.}
\end{acknowledgements}

\bibliographystyle{aa}
\bibliography{list_bib}
\begin{appendix}
\section{Wavelet artifact around A3667}
\label{sec:app_artifact}
The wavelet-filtered images show a wavelet artifact encircling A3667 (e.g., \cref{Fig:SystemOverview}, \cref{Fig:SB_boxes} top panel). These artifacts are horseshoe-shaped regions around elliptical sources where less emission is visible than there is actually present. The enhancement of emission northwest of A3667 could either be real or it could simply be the case that this part is not affected by the wavelet artifact and that there is in fact similar emission in the other regions around the cluster. To investigate this, we performed SB analysis around the cluster.

\begin{figure}[htbp]
        \centering
        \includegraphics[width=\hsize]{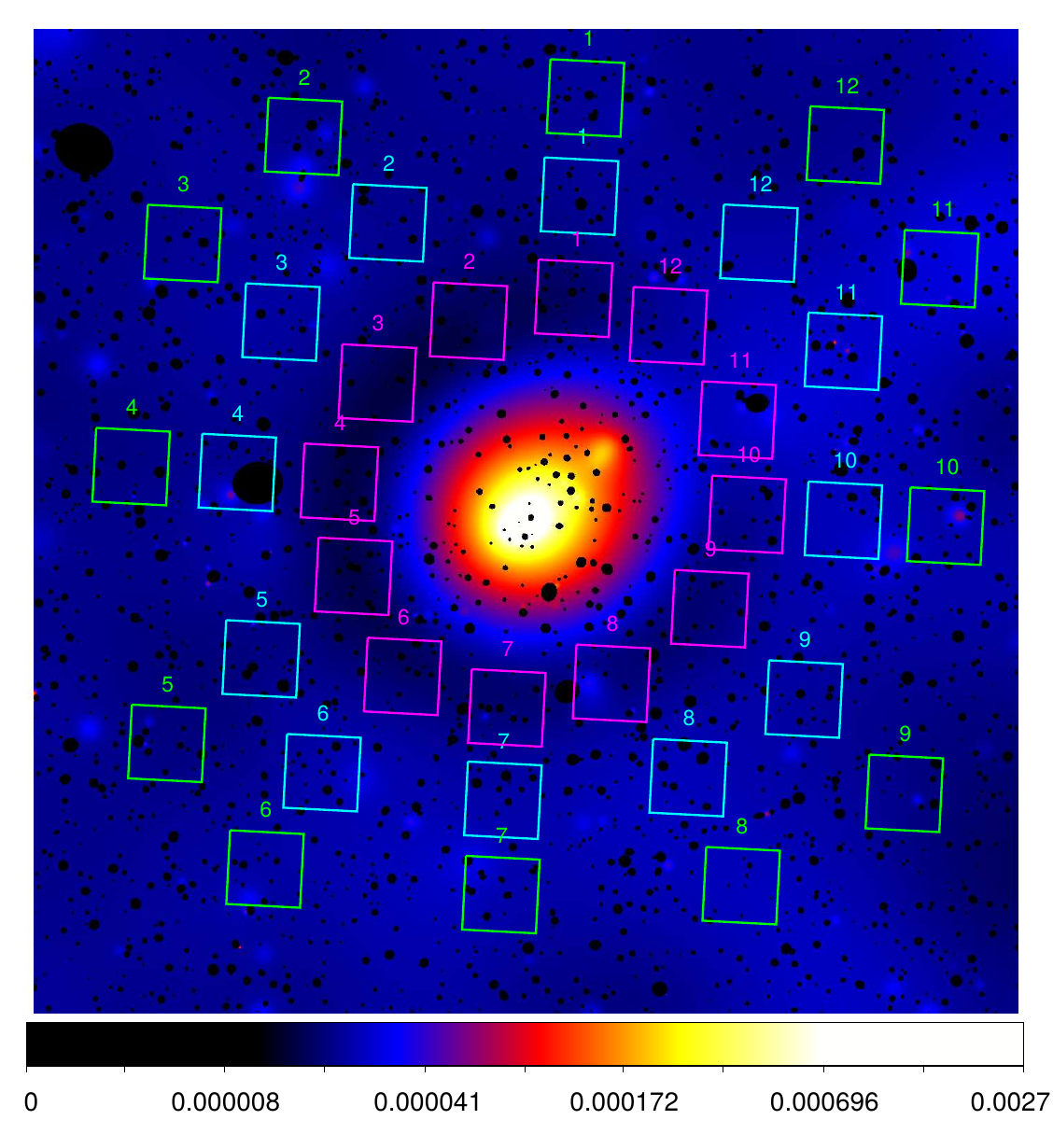}
        \includegraphics[width=\hsize]{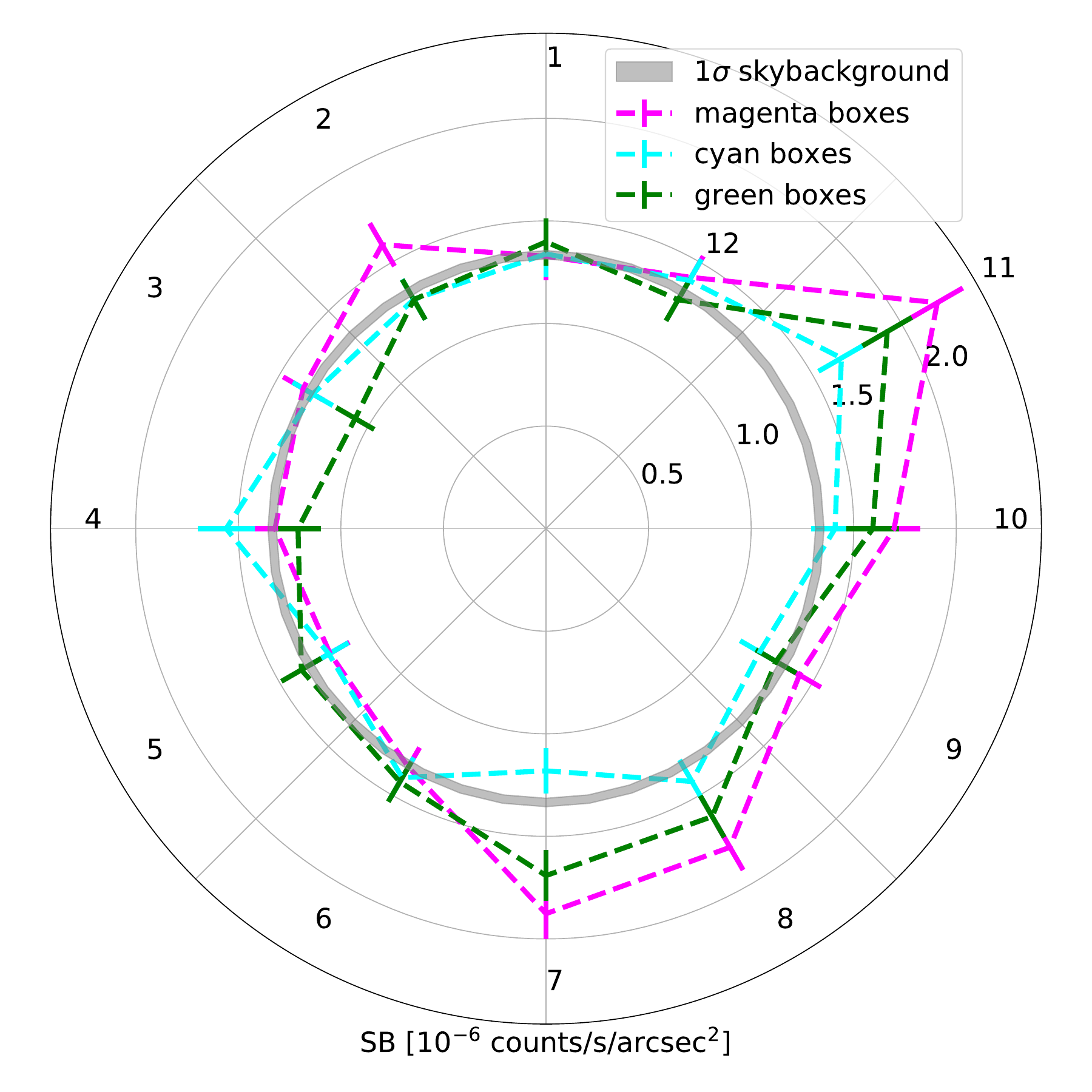}
        \caption{The wavelet artifact encircling A3667. Top: Data-reduced and wavelet-filtered image of A3667. Three rings with twelve boxes each are drawn to examine the horseshoe-shaped structure around the cluster. Bottom: SB values of the boxes in the left in a polar diagram. The dashed lines between the data points are only for the purpose of better visualization and have no physical meaning. The gray circle displays the SB of the sky background.}
        \label{fig:artifact_SB}
\end{figure}

In the top panel of \cref{fig:artifact_SB}, three rings, each consisting of twelve $\SI{0.25}{\degree}\times\SI{0.25}{\degree}$ boxes, are drawn around A3667 between ${\sim} R_{200}$ and ${\sim} 3R_{200}$. The size of the boxes was chosen such that a statistically representative number of photon counts are contained in each box, but also such that as many boxes as possible fit in the image in order to get a better spatially resolved result. For these boxes, the SB is calculated via \cref{eqn:SB_formula}.
In the bottom panel of \cref{fig:artifact_SB}, the calculated SB values for each box with their corresponding error bars are plotted into a polar plot. The polar plot is chosen in such a way that the numbers and colors of the boxes in the top panel correspond to the numbers and colors in the bottom panel. This allows the viewer to directly compare the structures in the two images. The displayed sky background is the weighted mean of the background boxes from \cref{Fig:SB_boxes}.
A comparison of the top and bottom panel of \cref{fig:artifact_SB} shows that indeed most of the horseshoe-shaped structure can be attributed to a wavelet artifact. According to the top panel, the emission in the boxes~1,~2,~3,~7,~9 in the magenta annulus should be considerably lower than in the other annuli, which, judging by the bottom panel, is definitely not the case. Nevertheless, there are also a few real structures: The emission in boxes~5~and~6 in the magenta annulus is in fact lower than in the other annuli and, even more important, the emission in box~11 is in fact stronger than in any other direction. Furthermore, except for boxes~7,~8,~10~and~11, the emission is not considerably higher than the sky background which is represented by a gray circle in the plot. So in summary, the horseshoe-shaped structure in the wavelet image is mainly a wavelet artifact, but it is also true that there is more emission in the direction of the filament (box~10~and~11).

\FloatBarrier
\section{S0840}
\label{sec:app_S0840}
\begin{figure}[htbp]
        \centering
        \includegraphics[width=\hsize]{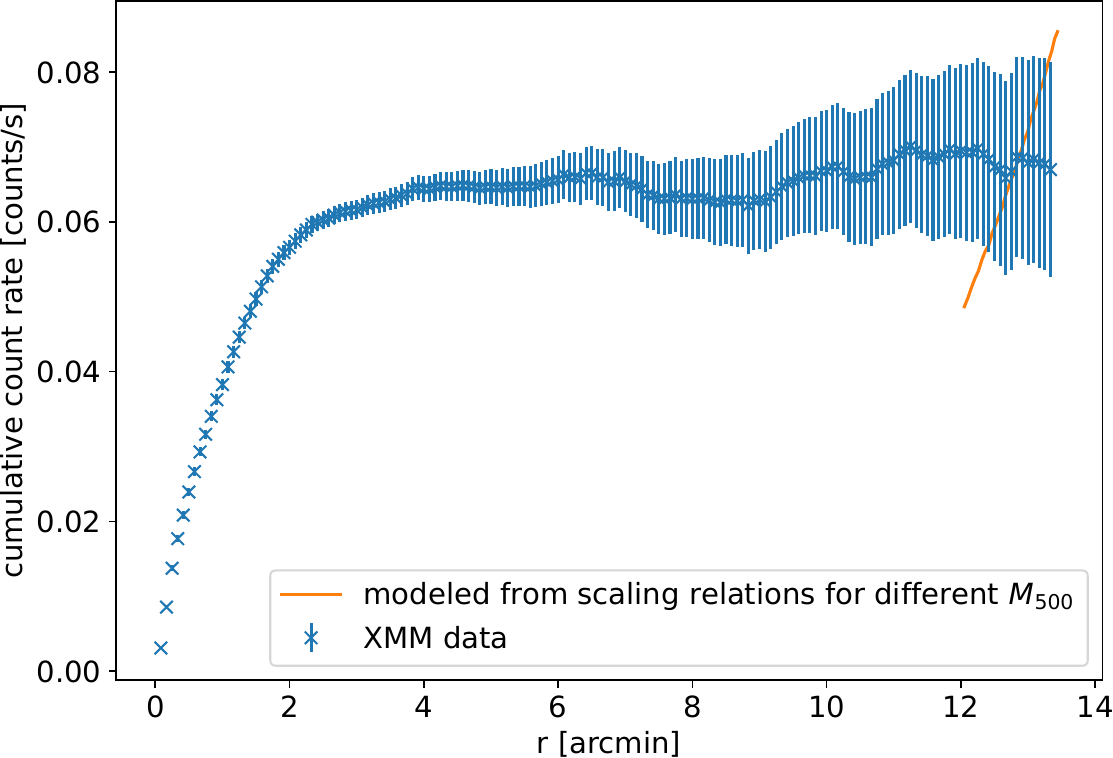}
        \caption{Cumulative count rate for Abell~S0840 and for modeled galaxy cluster groups with different masses. The intersection of both curves provides us with the $R_{500} = \SI{12.9(5)}{arcmin}$.}
        \label{Fig:growth_curve}
\end{figure}

In order to estimate the influence of Abell~S0840 on the filament flux, the $R_{500}$ of Abell~S0840 and its luminosity and flux inside $R_{500}$ were determined using scaling relations from \citet{Lovisari_2015}. Starting from an initial mass range, the expected cumulative count rate for a cluster with the given mass was modeled and plotted together with the cumulative count rate of Abell~S0840 in \cref{Fig:growth_curve}. The intersection of both curves provides us with the $R_{500} = \SI{12.9(5)}{arcmin}$. The uncertainty is estimated by taking the intersection of the modeled curve with the lower and upper uncertainties of the count rate. This new $R_{500}$ is $\sim\SI{20}{\percent}$ smaller than the $R_{500} = \SI{15.6}{arcmin}$ from MCXC \citep{MCXC}. The flux and luminosity inside the redetermined $R_{500}$ is $F_{500} = \SI{4.5(9)e-13}{erg\per\second\per\centi\meter\squared}$ and $L_{500} = \SI{2.2(3)e+41}{erg\per\second}$, respectively. For such a low-luminosity group, it is reasonable to trace the SB only in the central part, which is the reason we see significant emission to maximal $\sim\frac{1}{3}$ of the $R_{500}$. And the projected size of the $R_{500}$ is only comparable to the ones of the other two clusters because the redshift is $\sim\num{4}$ times lower.
The leftover flux of Abell~S0840 beyond the $\SI{4}{arcmin}$ excised circle in the SB analysis of the filament is  $F_{\textnormal{leftover}} = \SI{3(8)e-14}{erg\per\second\per\centi\meter\squared}$.

\FloatBarrier
\section{Spatial and redshift distribution of galaxies in the whole FoV}
\label{sec:app_zdistri}
\begin{figure}[htbp]
        \centering
        \includegraphics[width=\hsize]{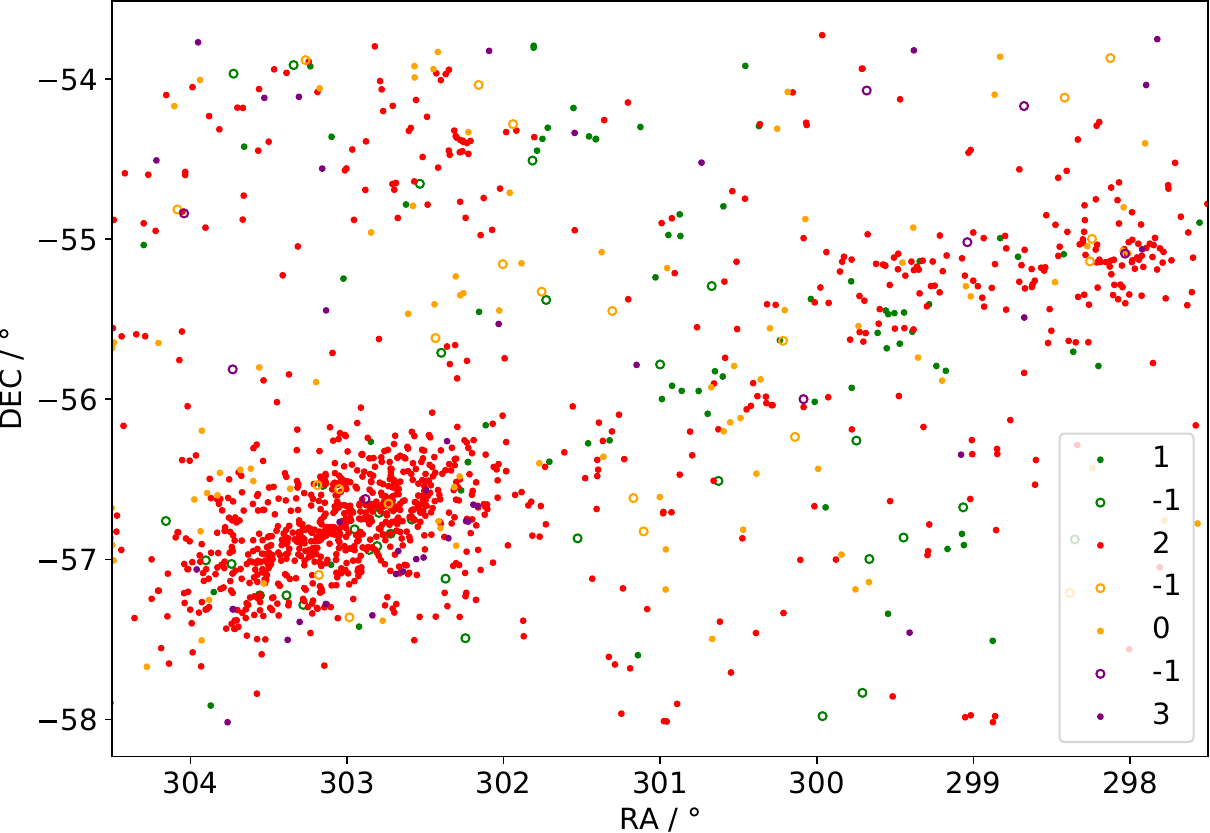}
        \includegraphics[width=\hsize]{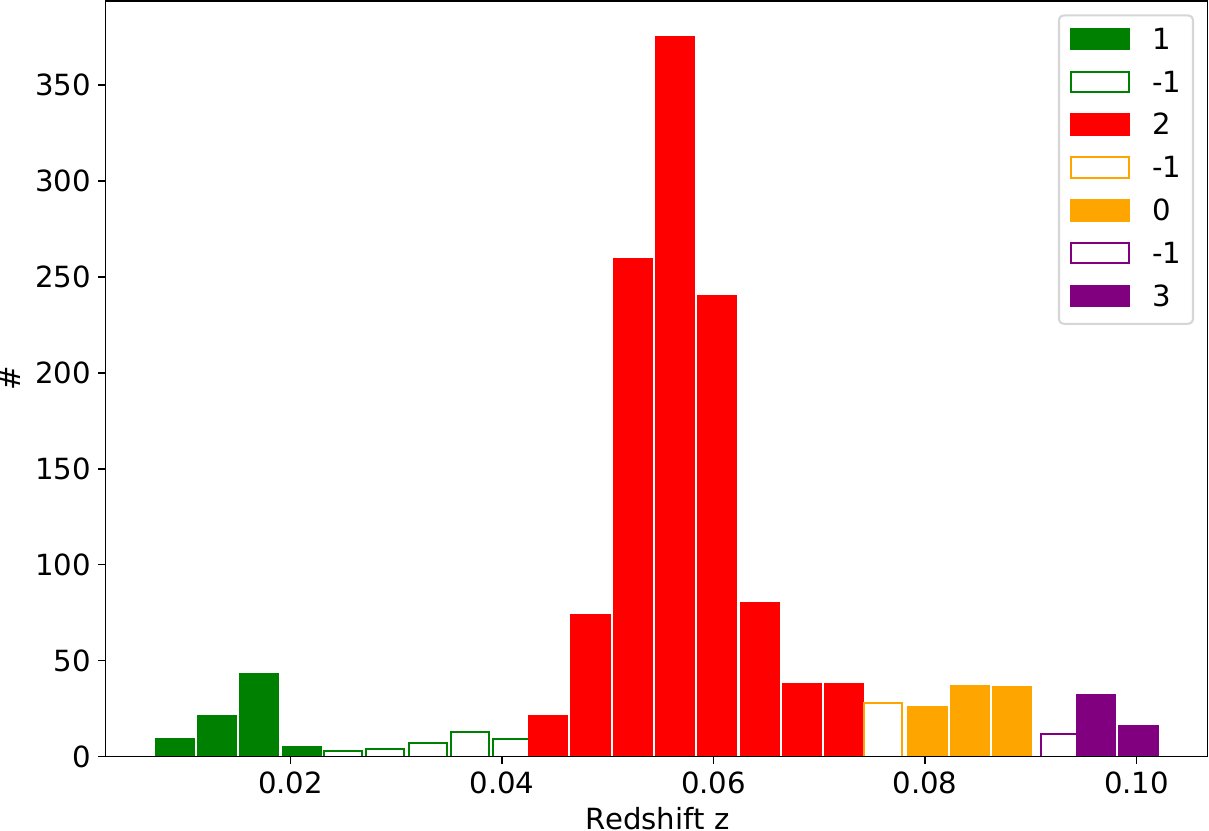}
        \caption{Spatial (top) and redshift (bottom) distribution of galaxies in the whole FoV.    
    By comparing the plots in the top and bottom panel it can be seen where galaxies with certain redshifts are located spatially. For information about the colors and labels see Sect. \ref{sec:z_analysis} and \cref{sec:app_zdistri}.}
        \label{fig:FoV_galaxies}
\end{figure}

The analysis in \cref{Fig:DBScan_plot} is only for the galaxies inside circles~2-4, that is, the region of interest for the filament. To give a more comprehensive overview of the whole FoV the same analysis is done for all NED galaxies in the FoV with $z < 0.1$ in \cref{fig:FoV_galaxies}. The colors and labels are showing the same redshift categories as in \cref{Fig:DBScan_plot}. Only the redshift intervals where previously no galaxies, or only outliers that could not be identified with a redshift category, are located, are colored differently. These intervals still have a label of $-1$ and have the same color as a nearby redshift interval, but they are shown with edgecolors instead of facecolors. This way it is possible to identify galaxies with similar redshifts by color and the information of the the redshift cuts from DBSCAN is preserved.

The analysis shows that the vast majority of the galaxies (red) are still close-by galaxies, which is expected since almost all of them are member galaxies of the rich galaxy cluster A3667. The regions with higher galaxy density seen in the 2MASS map (\cref{Fig:2MASS_both} left panel) can also be identified here. Furthermore, the plots show that galaxies of a specific redshift are not restricted to a specific spatial region but are spread over the complete FoV. This is in particular of importance for the red galaxies since it shows that the overdensity of close-by galaxies is not caused by the fact that previous analyses and redshift surveys could have been performed especially in the regions of the galaxy clusters. It is also interesting for the low redshift galaxies (green) to see that these are not restricted to the position of S0840 but similar patterns can be found in the FoV, further reducing the impact of this galaxy group on the filament analysis.

\FloatBarrier
\section{Spectral analysis fitting parameters and robustness tests}
\label{sec:app_spectral}
\change{\cref{tab:spectral_results} shows the resulting fit parameters for the source region using the two different background estimates. The area of the source region is $\SI{4462.17}{arcmin\squared}$ (without the area of the excised sources). The results for the temperatures agree within  $\SI{1}{\sigma}$, the metallicities differ by $\SI{1.2}{\sigma}$.}

\change{The same fitting was performed multiple times with the following modifications:
\begin{itemize}
    \item using $\SI{2.3}{keV}$ instead of $\SI{9.0}{keV}$ for the upper limit of the energy range,
    \item using the redshift of A3667 or the redshift of A3651 or the mean of these two redshifts for the filament redshift,
    \item using only neutral $N_{\text{HI}}$ from the HI4PI all-sky survey \citep{HI4PI} instead of $N_{\text{H,tot}}$ by \citet{Willingale_13},
    \item using the minimum and maximum $N_{\text{H}}$ of the spectrum extraction box (tested for source box as well as background box) instead of using the value at the central position. These values differ by $\approx\SI{10}{\percent}$.
\end{itemize}
The results for the first three tests change only by less than $\SI{0.3}{\sigma}$. For the last test the results change by less than $\SI{0.8}{\sigma}$.}

\begin{table}[htbp]
   \caption{\change{Results from spectral fitting using both background regions}}
   \label{tab:spectral_results}
   \renewcommand{\arraystretch}{1.5}
   $$
   \begin{array}{lll}
      \hline
      \noalign{\smallskip}
      \textnormal{Parameter} & \textnormal{Background North} & \textnormal{Background South} \\
        & N_{\text{H,tot}} = \num{5.1} \ ^{\S} & N_{\text{H,tot}} = \num{4.5} \ ^{\S} \\
      \noalign{\smallskip}
      \hline
      \noalign{\smallskip}
      T \text{ / keV} & 0.86^{+0.08}_{-0.06} & 0.91^{+0.07}_{-0.05} \\
      Z \text{ / } Z_\odot & 0.04^{+0.03}_{-0.02} & 0.10^{+0.05}_{-0.04} \\
      \text{norm source }^{\dagger} & 2.5^{+0.5}_{-0.4} & 1.7^{+0.4}_{-0.3} \\
      \text{norm LHB }^{\dagger} & \num{2.7(2)} & \num{3.3(2)} \\
      \text{norm MWH }^{\dagger} & \num{1.98(5)} & 1.95^{+0.05}_{-0.04} \\
      \text{norm CXB }^{\ddagger} & \num{7.0(2)} & \num{7.1(2)} \\
      \noalign{\smallskip}
      \hline
   \end{array}
   $$ 
   {\raggedright $\S$ in $\num{1e+20}$\,cm$^{-2}$ \\ $\dagger$ in $\num{1e-6}$\,cm$^{-5}$\,arcmin$^{-2}$ \\ $\ddagger$ in $\num{1e-7}$\,photons\,keV$^{-1}$\,s$^{-1}$\,cm$^{2}$\,arcmin$^{-2}$ at 1\,keV\par}
\end{table}

\end{appendix}

\end{document}